\documentclass{aa} 
\usepackage{longtable}
\usepackage{graphicx}
\usepackage{txfonts}
\usepackage{lipsum}
\usepackage{caption}
\usepackage{subcaption}     % necessary for continued figures, example in section 3
\usepackage{lscape}       % to rotate a single page table, example in appendix.
\usepackage{placeins}      % useful with \FloatBarrier, to keep 
\usepackage{hyperref}
\usepackage{booktabs}
\usepackage{multirow}
\newcommand {\um} {\,{\mu\rm m}}
\newcommand{\Myr}{\,{\rm Myr}}
\newcommand{\bagpipes}{\textsc{Bagpipes}}

\begin{document}

%%%%%%%%%%%%%%%%%%%%%%%%%%%%%%%%%%%%%%%%
% if you use custom commands in your title,
% ensure to check your title when submitting!
%%%%%%%%%%%%%%%%%%%%%%%%%%%%%%%%%%%%%%%%
  
\title{An HI+radio continuum study of local cluster galaxies:}
\subtitle{Intercepting the early stages of environmental processing with MeerKAT}
  
%%%%%%%%%%%%%%%%%%%%%%%%%%%%%%%%%%%%%%%%
% Please separate each author with the \and command
%
% Please do not include ORCIDs next to author names.
% Only ORCIDs authenticated by individual authors in EDPS
% editorial system will be taken into account.
% ORCIDs included here will be removed.
%%%%%%%%%%%%%%%%%%%%%%%%%%%%%%%%%%%%%%%%

  \author{
  Alessandro Ignesti\inst{1,2}
  \and
  Paolo Serra\inst{3}
  \and
  Alessandro Bianchetti\inst{1}
  \and
  Antonino Marasco\inst{1}
  \and
  Alessia Moretti\inst{1}
  \and
  Francesca Loi\inst{3}
  \and
  Bianca M. Poggianti\inst{1}
  \and
  Benedetta Vulcani\inst{1}
  \and
  Laura Bisigello\inst{1}
  \and
  Myriam Gitti\inst{4,5} 
  \and
  Marco Gullieuszik\inst{1}
  \and
  Augusto E. Lassen\inst{1}
  \and
  Rosita Paladino\inst{5,6}
  \and
  Christoph Pfrommer\inst{7}
  \and
  Mpati Ramatsoku\inst{8,3}
  \and
  Francesco Sinigaglia\inst{9,10,11,12}
  \and
  Neven Tomi\v{c}i\'{c}\inst{13}
}

\institute{
  INAF - Osservatorio Astronomico di Padova, Vicolo dell'Osservatorio 5, 35122 Padova (PD), Italy \label{inst1}
  \and
  Astronomical Institute of the Czech Academy of Sciences, Bo\v cn\'i II 1401, 14100 Prague, Czech Republic\label{inst2}\\\email{alessandro.ignesti@asu.cas.cz}
  \and
  INAF - Osservatorio Astronomico di Cagliari, Via della Scienza 5, 09047 Selargius (CA), Italy \label{inst3}
   \and
  Department of Physics and Astronomy "Augusto Righi", University of Bologna, Via Gobetti 93/2, 40129 Bologna (BO), Italy \label{inst4}
  \and
  INAF - Istituto di Radioastronomia, Via Piero Gobetti 101, 40129 Bologna (BO), Italy \label{inst5}
  \and
   ESO, Karl Schwarzchild Str. 2, D-85478 Garching bei München, Germany \label{inst6}
   \and
  Leibniz Institute for Astrophysics Potsdam (AIP), An der Sternwarte 16, 14482 Potsdam, Germany \label{inst7}
  \and
  Centre for Radio Astronomy Techniques and Technologies (RATT), Department of Physics and Electronics, Rhodes University,
Makhanda 6140, South Africa \label{inst8}
\and
  Institute for Fundamental Physics of the Universe, Via Beirut 2, I-34151 Trieste, Italy \label{inst9}
  \and
  SISSA - International School for Advanced Studies, Via Bonomea 265, 34136 Trieste, Italy \label{inst10}
  \and
  INAF - Osservatorio Astronomico di Trieste, Via G. B. Tiepolo 11, I-34131 Trieste, Italy \label{inst11}
  \and
  INFN – National Institute for Nuclear Physics, Via Valerio 2, I-34127 Trieste, Italy \label{inst12}
   \and
    Department of Physics, Faculty of Science, University of Zagreb, Bijeni\v{c}ka 32, 10 000 Zagreb, Croatia \label{inst13}
}
\authorrunning{Ignesti et al.}

  \date{Received September 30, 20XX}

% \abstract{}{}{}{}{}
% 5 {} token are mandatory
 
 \abstract
% {The last dance}
{The evolution of galaxies in clusters is driven by their interaction with the environment, which deeply affects and alters the properties of the multi-phase interstellar medium. Here we make use of MeerKAT observations to study the properties of the neutral and nonthermal interstellar medium, traced respectively by the HI 21 cm line and by the radio continuum emission at 1.4 GHz, out to about twice the virial radius of three local ($z\simeq0.04$) galaxy clusters, namely IIZW108, A4059 and A3558, with $M_{200}\simeq2-9\times10^{14}~M_\odot$. We assemble a sample of 61 galaxies, including the so-called jellyfish and unwinding galaxies, detected in both HI and radio continuum, and derive their optical properties from IR-to-UV photometry using ancillary data. We find that cluster galaxies in our sample are on average $\sim2.7$ times more star-forming than galaxies with the same stellar and HI mass in the field, indicating that the HI + radio-continuum selection has intercepted galaxies at the very early stages of their environmental processing when the external pressure has not yet removed the gas, but the resulting fast compression has enhanced both the star formation and the radio continuum luminosity. The study also reveals that galaxies at the initial stage can features a radio luminosity excess due to the old relativistic electrons permeating the interstellar medium. We show that unwinding galaxies are characterized by high stellar and HI masses, arguing that their peculiar morphology may results from their large baryonic mass. Finally, via the stacking analysis of 677 optically-selected cluster members we quantitatively show that cluster galaxies are more HI-poor than in the field, and the deficiency steadily grows approaching the cluster center where galaxies have, on average, a factor $\sim3\times$ less HI mass than those in the field. }
 % {Mandatory. The objectives of the paper are defined here.} 
  %{Mandatory. The methods of the investigation are outlined here}
  %{Mandatory. The results are summarized here.}
  %{Optional, leave empty if necessary. “Conclusions” can be used to explicit the general conclusions that can be drawn from the paper.}


  \maketitle

%%%%%%%%%%%%%%%%%%%%%%%%%%%%%%%%%%%%%%%%%%%%%%%%%%%%%%%%%%%%%%
\nolinenumbers
\section{Introduction}

The balance between the gas inflow and removal drives the baryon cycles in galaxies, and thus their evolution. Whereas the inflows sustain star formation and allow galaxies to grow in mass, gas removal can lead to star formation quenching. Galaxies can lose their gas via internal or external processes where the latter, generally known as environmental effects, are crucial in shaping galaxy properties in clusters and groups \citep[][]{Dressler1980,Cortese_2021,Vulcani_2022,Boselli_2022}.
They can be divided into two main categories, those driven by gravitational forces, such as mergers and tidal interaction \citep[][]{Barnes_1992} or harassment due to fast encounters \citep[][]{Moore1996}, and those resulting from the hydrodynamical interaction between the galaxies and the environmental plasma, either the intracluster medium (ICM) or the intragroup medium. This category includes thermal evaporation \citep[][]{McKee-Cowie_1977}, viscous stripping \citep[][]{Nulsen1986}, and ram pressure stripping \citep[RPS,][]{Gunn1972}. The latter is considered to be ubiquitous in clusters, supported by evidence that every galaxy goes through at least one ram pressure stripping event during their staying in the cluster \citep[e.g.,][and references therein]{Vulcani_2026}, and the evidence, collected with HI observations, that the more cluster galaxies approach the cluster center, the more they are HI-poor with respect to the field \citep{Giovannelli_1985, Gavazzi2006,Chung_2007, Deb_2023, Reynolds_2022}. Ram pressure is caused by the external pressure exerted by the environmental plasma on a moving body, typically expressed as $P_{\text{ram}}=\rho V^2$ where $\rho$ is the medium density and $V$ is the velocity difference between the object and the medium \citep[][]{Gunn1972}. In the most extreme cases, ram pressure can result in dramatic gas loss for a galaxy, occurring when the pressure exerted by the hot ICM on the interstellar medium (ISM) overcomes the gravitational anchoring force. The gas loss induced by RPS can effectively quench star formation in the stellar disk \citep[][for a review]{Cortese_2021,Boselli_2022}, thus making it an important quenching pathway for cluster galaxies \citep[e.g.,][]{Vollmer_2001,Tonnesen_2007,Vulcani2020,Watts_2023,Marasco_2026,Vulcani_2026}. Furthermore, environmental processing can be preceded by the so-called "pre-processing" in group environments or along cosmic web filaments \citep{Vulcani_2021,Vulcani_2026}, as lower relative velocities in these regimes facilitate efficient gravitational interactions and early-stage gas losses \citep{For_2021, Yoon_2025}.

Optical data are powerful probe of ram pressure physics as they allow us to study the impact of RPS on the star formation properties and on the overall galaxy morphology, and can ultimately provide a view of the "long-term" effects of the environment in cluster galaxies. Recent studies based on integral field spectroscopy observations have shown how the hydrodynamic compression of residual gas during the initial stages of RPS can temporarily boost star formation efficiency before the total depletion of the gaseous reservoir \citep{Vulcani_2018c}. Furthermore, the ram pressure action is not limited to displacing the ISM outside of the disk. The impact of ram pressure can result in many effects, including compression of gas along the leading edge of the disk \citep[e.g.,][]{Rasmussen_2006,Poggianti_2019,Roberts_2022e}, disturbed galaxy morphologies, trailing tails of stripped gas \citep[e.g.,][]{Kenney2004,vanGorkom_2004,Fumagalli2014,Poggianti2017}, and condensation of star-forming knots in the tails \citep[][]{Kenney2014,Poggianti2019}. It can also temporarily enhance the global star formation \citep[e.g.,][]{Poggianti2016,Vulcani2018,Roberts_2020, Ramatsoku_2020} and trigger the activity of the central nuclei \citep[e.g.,][]{Poggianti2017b,Peluso_2021}. Ram pressure can also affect the microphysics of the ISM on small scales, for example by stimulating the conversion from atomic to molecular hydrogen \citep[][]{Moretti_2020}, enhancing the diffusivity of cosmic rays \citep[][]{Farber_2022,Ignesti_2022b}, or by inducing mixing between the ISM, circumgalactic medium (CGM), and ICM \citep[][]{Franchetto_2021,Sun_2022,Sparre2024}.

Given its broad impact on galaxy physics, it would be especially important to study galaxies at the different stages or RPS to understand the series of transformations that leads to the galaxy quenching. Optical morphology can help selecting galaxies under ram pressure, as RPS can thereby induce a drastic transformation in a galaxy morphology \citep[][]{Lassen2026}, such as the 'unwinding' of their spiral arms \citep[][]{Bellhouse_2021}, and lead to developing trails of stripped ISM material. The most extreme examples of galaxies undergoing strong RPS are the so-called jellyfish galaxies \citep{Fumagalli2014, Smith2010, Ebeling2014,Poggianti2017} which show extraplanar, unilateral debris extending beyond their stellar disks, and striking tails of ionised gas hosting star-forming regions. However, optical data are  poor tracer of the very early stage of the process, that is when the ISM begins to be directly stripped. Tracing this stage is particularly complicated as its time-scale is expected to be relatively short \citep[$<0.5$ Gyr,][]{Marasco_2026}, but is a fundamental piece of the puzzle. Hence radio observation of the  neutral atomic hydrogen (HI) gas can represent a more powerful tracer, because RPS constitutes a primary mechanism for HI removal \citep{Poggianti_2001,Chung_2007, Wang_2021}. Morphologically, environmental processes are evidenced by one-sided HI tails, extra-planar emission, and significantly truncated gas disks \citep{Chung_2007, Loni_2021, Wang_2021, cakir_2026}. Radio observations can also probe on the nonthermal ISM, i.e. magnetic field and cosmic rays that we can trace thanks to their synchrotron continuum emission \citep[][]{Condon_1992,Beck_2000}. Radio continuum observations have revealed that RPS can leave a series of characteristic signatures in star-forming galaxies. RPS galaxies are typically in excess of radio luminosity with respect to their ongoing star formation rate \citep[][]{Murphy2009,Vollmer_2013,Chen_2020, Roberts_2021a,Ignesti_2021,Ignesti_2022b,Edler_2024} and show characteristic radio continuum tails, produced by the electrons accelerated in the stellar disk and subsequently stripped by the ram pressure \citep[e.g.,][]{Vollmer_2004,Chen_2020,Muller_2021,Ignesti_2021,Roberts_2021c,Venturi_2022, Ignesti_2023}. 
Finally, combined observational and theoretical studies have shown that RPS galaxies can interact with the ICM magnetic field through the so-called magnetic draping \citep[][]{Dursi_2008,Pfrommer_2010, Ruszkowski_2014,Muller_2021, Sparre2024,Ignesti_2026}. 

In this work, we present new MeerKAT observations of three low-redshift clusters, IIZW108, A4059 and A3558 to explore how the environment affect their galaxies. The three clusters cover a broad range in mass ($M_{200}=2-9\times10^{14}~M_\odot$, see Table \ref{table_data}) and dynamical status, with A3558 being a merging system residing in the central region of the Shapley Supercluster \citep[][]{Shapley_1930,Rossetti_2007,GdG_2025}, and A4059 being a relaxed cool-core cluster \citep[e.g.,][]{Reynolds_2008,Mernier2015}. Here we make use of the full potential of the MeerKAT telescope which, by observing the full bandwidth and with the maximum frequency resolution can simultaneously observe both the neutral and the nonthermal ISM emission, to perform the first HI+radio continuum study of the combined galaxy population of the three clusters. The manuscript is structured as follows. The MeerKAT data processing and ancillary data preparation are presented in Section \ref{section1}. Results are presented in Section \ref{section2} and further discussed in Section \ref{section3}. Throughout the paper, we adopt a $\Lambda$CDM cosmology with $\Omega_{\Lambda}=0.7$, $\Omega_{\rm m}=0.3$, and $H_0=70$ km s$^{-1}$ Mpc$^{-1}$. For the clusters analyzed in this work, it results in $1''\simeq0.9$ kpc.% through the paper we adopt the initial mass function presented in \citet{Chabrier2003}.

\section{Data reduction}
\label{section1}
\subsection{MeerKAT data analyisis}
\begin{table}[]
\caption{\label{table_data}Galaxy cluster parameters.}
\setlength{\tabcolsep}{5pt} % Default value: 6pt
\renewcommand{\arraystretch}{1.5} % Default value: 1
  \centering
  \begin{tabular}{l|ccc}
  \toprule
  \toprule
     & IIZW108 & A4059 & A3558 \\ 
     \midrule
  Coord. & \begin{tabular}{@{}c@{}}318.483 \\ +2.5654\end{tabular} &\begin{tabular}{@{}c@{}}359.2535 \\ -34.7599\end{tabular}&\begin{tabular}{@{}c@{}}201.9891 \\ -31.496\end{tabular}\\
  \midrule
  $z_\text{cl}$&0.04889&0.04877&0.04829\\
  \midrule
 M$_{200}$ [$10^{14}~M_\odot$] &2.05&4.70&8.83\\
 \midrule
  $\sigma$ [km s$^{-1}$]&575&744&910\\
  \midrule
  HI band [MHz]& \begin{tabular}{@{}c@{}}1342.79-\\1359.13 \\ (16.3)\end{tabular} &\begin{tabular}{@{}c@{}}1340.56-\\1361.7\\ (21.1)\end{tabular} & \begin{tabular}{@{}c@{}}1338.89-\\1364.74\\ (25.9)\end{tabular} \\
  \midrule
  $\Delta V$ [km s$^{-1}$]& \begin{tabular}{@{}c@{}}12258-\\15708\end{tabular} &\begin{tabular}{@{}c@{}}12399-\\16863\end{tabular} & \begin{tabular}{@{}c@{}}11757-\\17217\end{tabular} \\
  \midrule
  Calibrators &\begin{tabular}{@{}c@{}c@{}}J1939-6342 \\ J2134-0153 \\ J0108+0134\end{tabular}&\begin{tabular}{@{}c@{}c@{}}J1939-6342 \\ J1323-4452 \\ J1331+3030\end{tabular}&\begin{tabular}{@{}c@{}c@{}}J1939-6342 \\ J1323-4452 \\ J1331+3030\end{tabular}\\
  \bottomrule
  \end{tabular}
   \tablefoot{From top to bottom: Galaxy cluster name; J2000 cluster center sky coordinates; Redshift, virial mass and velocity dispersion from \citet[][]{Gullieuszik2020}; Frequency band (bandwidth) corresponding to $z_{\text{cl}}\pm3\sigma$ used to detect HI sources and corresponding optical velocity coverage $\Delta V$; Amplitude, phase and polarization calibrators adopted in the observations.}
\end{table}

The three galaxy clusters were observed with MeerKAT under the project SCI-20220822-AM-03 (PI Müller). Broadband full-Stokes data were acquired between December 2022 and January 2023 using the 32k correlator. Each target has been observed for a total of eight hours, divided into two four-hours runs, in L band (856-1712 MHz) in full stokes, and with the maximum frequency resolution of 26.123 kHz, corresponding to 32768 channels.  For each target, a suited sets of calibrators (see Table \ref{table_data}) have been observed. Each run included a 10-minute scan of a primary calibrator. A secondary calibrator was observed for 2 minutes both before and after the target scans. The target sources have been observed for 30 minutes per scan, accumulating a total of 4 hours of on-source integration time per observing session. Additionally, two 5-minute scans of the polarized calibrators have been conducted at different parallactic angles to optimize sensitivity in the cross-hand correlations, ensuring adequate signal strength for polarization calibration. The total amount of raw data collected is 26.81 Tb.

In order to investigate both the continuum and the HI emissions, each run has been subsequently split into two different sub-datasets, for a total of twelve datasets. We created  a high-resolution one to study the HI emitters within the clusters, centered on the clusters' frequency and spanning three times their velocity dispersion (see Table \ref{table_data}) processed at the native channel resolution of 26.123 kHz ($\sim5.5$ km s$^{-1}$ at $z=0$), and a low-resolution one with a channel width of 208 kHz covering the whole band to study the continuum sources.

The data reduction was carried out at the INAF-OAC computational center on the \texttt{Escondida2} mini-cluster equipped with 1 TB of RAM and 2 CPUs, for a total of 56 threads, and by using the \texttt{CARACal} pipeline \citep[][]{jozsa2020}. The data reduction follows the procedures outlined in \citet[][]{Serra_2023} for the HI and \citet[][]{Loi_2025} for the continuum. After splitting the calibrators in new measurement sets, we flag autocorrelations, shadowed antennas, and the frequency ranges 1379.6--1382.3 MHz and 1420.36--1420.56 MHz affected by the GPS L3 signal and by absorption/emission of neutral hydrogen from the Milky Way respectively. The \texttt{AOFlagger} \citep[][]{offringa2012} tool has been used to excise the remaining RFIs using the QUV Stokes visibilities. 
After deriving the calibration solutions as described in the aforementioned works, we applied them to the target, repeated the flagging steps and proceed with the self-calibration loop. For this purpose, we performed two rounds of self-calibration. For the high-resolution data we derived new phase solutions every 320 seconds, and for the low-resolution sub-data sets we adopted a phase-and-delay solution interval of 32 seconds.

As a caveat, we report that our daytime observations, during which the distance from the sun ranged from 39 to 146 degrees, resulted in severe solar RFI. This caused a significant data loss of $\sim30\%$ in the HI datasets and up to $60\%$ for the continuum. In particular, one of the A3558 runs, which also has the lowest angular separation from the Sun and was performed partly during sunrise, exhibited extreme solar RFIs which forced us to ultimately drop it from the analysis.

For the preparation of the final continuum images, we further masked the bandwidth corresponding to the HI window (see Table \ref{table_data}) to avoid contamination from the line emission. The imaging has been carried out with the \texttt{WSCLEAN} software \citep[][]{Offringa_2014} cleaning within regions defined by the \texttt{SoFia-2} tool \citep[][]{serra2015,Westmeier2021MNRAS.506.3962W}. Radio continuum images have a resolution of 20 arcsec. Due to the lack of uniform H$\alpha$ coverage, we did not correct for the thermal components but we note that, at this frequencies, it should account for less than $6\%$ of the total continuum emission \citep{Ignesti_2022b}. The final RMS level of the continuum images for IIZW108, A4059, and A3558 are, respectively, 7.2, 8.0 and 4.2 $\mu$Jy beam$^{-1}$. Radio continuum images have been corrected for the primary beam by following \citet{Mauch_2020}.

We made the HI cubes as follows. First, we subtracted the continuum model obtained during self-calibration from the visibilities. We then generated line cubes with \texttt{WSCLEAN} applying a UV tapering of 30 arcsec  in order to maximise the sensitivity to extended, faint emission. This resulted in a resolution of $\sim40$ arcsec, corresponding to $\sim38$ kpc at the clusters' distances. The HI cubes were initially cleaned blindly by \texttt{WSCLEAN}. We then created an HI detection mask with \texttt{SoFia-2}. We remove residual continuum emission and artifacts, either caused by incomplete sky model and/or imperfect gain calibration, by subtracting the best-fitting spline from each sightline of the HI cubes with \texttt{imcontsub}\footnote{\url{https://github.com/laduma-dev/mowjsub}}. When doing this, voxels in the HI detection mask were excluded from the fit. We then run \texttt{SoFia-2} on the improved HI cubes to create a final mask. In order, the final mask was used 1) as a clean mask to make the HI cubes again with \texttt{WSCLEAN}, 2) as an exclusion mask to fit and subtract the residual continuum emission from the new cubes with \texttt{imcontsub}, and 3) as a detection mask to make HI moment maps and single-detection HI products from the continuum-subtracted cubes with \texttt{SoFia-2}.

We achieved HI column density limit of $(2.7-14.7)\times10^{19}$ cm$^{-2}$ at the 3$\sigma$ level for a linewidth of 50-200 km s$^{-1}$, while the point source HI mass detection limit is log$_{10}M_\text{HI}/M_\odot$=8.37-8.68. 
The total number of HI detections is 288, which includes a large fraction of solar RFIs and artifacts that present sizes and flux density level comparable with those of faint cluster sources. Therefore, to disentangle the emission associated with cluster galaxies from RFIs and noise peaks, we performed a visual inspection by comparing the HI total intensity images with DESI Legacy \citep{Dey+19} DR10 R-band optical images by checking if the HI intensity peak overlaps with optical galaxies. We identified 116 sources associated with optical galaxies in total, corresponding to 40$\%$ of the HI detections. For these sources, we measured HI flux density and corresponding HI mass as described in \citet{Meyer_2017}, and the HI asymmetry parameter for the emission above 3$\sigma$, $A_\text{HI}$, following the $A_\text{mod}$ estimate method presented by \citet{Lelli_2014}. We then used the continuum images to measure the radio continuum luminosity corresponding to each source. First we determined the local RMS within 7 arcminute from the galaxy, then we computed the flux density and converted it in radio continuum luminosity at 1.4 GHz, $L_\text{R}$, for the emission above the 3$\sigma$ level. The typical RMS level varies from $6~\mu$Jy beam$^{-1}$ up to $20~\mu$Jy beam$^{-1}$ toward the primary beam edge or in the proximity of bright radio sources. We also compute the radio continuum asymmetry, $A_\text{RC}$, for the emission above the 3$\sigma$ level. 

\subsection{Photometry, stellar masses and SFRs}
Stellar masses ($M_\star$) and star formation rates (SFRs) for all galaxies in the HI-selected sample are determined following a procedure similar to that adopted in \citet{Marasco+25a}, based on the modeling of their broad-band spectral energy distribution (SED).
We summarise the procedure below, indicating the main differences with respect to the original approach of \citet{Marasco+25a}, to which we refer for additional details and for an in-depth discussion of the uncertainties.

We first collected image cutouts, with sizes of $3'\times3'$ and centered around the HI intensity-weighted centres, in the following broad-band filters: near- and far-ultraviolet ($NUV$ and $FUV$) from the Galaxy Evolution Explorer \citep[\emph{GALEX;}][]{GildePaz+07}; \emph{g}, \emph{r}, and \emph{z} from the DESI Legacy Imaging Surveys \citep{Dey+19}; $J$, $H$, and $K_{\rm s}$ from the Two Micron All Sky Survey \citep[2MASS;][]{Skrutskie+06}; $3.6\um$ and $24\um$ from the IRAC and MIPS cameras of the \emph{Spitzer Space Telescope} \citep{Werner+04}; and $3.4\um$ and $22\um$ from the Wide-field Infrared Survey Explorer \citep[WISE;][]{Wright+10}.
The association between galaxies and HI detections is based on the UV- and $g$-band images and, with the exception of a few flagged cases, is unambiguous. 

Our photometric measurements are based on adaptive elliptical apertures, using sizes, inclinations and position angles determined as in Appendix A of \citet{Marasco+25a}. However, in contrast with \citet{Marasco+25a}, here for each galaxy we determined a single aperture based on the $b$-band image and kept it fixed for all the other bands. As in the \texttt{SExtractor} software \citep[][]{Bertin1996}, the aperture properties are based on the image 2nd-order moments, providing typical aperture diameters ranging from 20\arcsec to 70\arcsec, with a mean of $\sim$40\arcsec. These apertures are substantially wider than the PSF of the poorest resolution images (12\arcsec of FWHM for WISE4), and we verified visually that they extend well beyond the galaxy region, ensuring an unbiased estimate of the total flux in all bands. This approach maximises the S/N for faint detections as those provided by 2MASS and \emph{GALEX} at these redshifts. Finally, all measurements are corrected for foreground extinction in the Milky Way, using the reddening maps of \citet{SchlaflyFinkbeiner11} and the extinction curve of \citet{Fitzpatrick99} for $R_{\rm V}\!=\!3.1$.

The SED modelling was done via the Bayesian Analysis of Galaxies for Physical Inference and Parameter EStimation (\bagpipes) code \citep{Carnall+18}, using the 2016 version of the \citet{BruzualCharlot03} stellar population synthesis models, and the initial mass function (IMF) of \citet{KroupaBoily02}.
As in \citet{Marasco+25a}, we adopted a non-parametric star formation history (SFH) in $7$ age bins ($0$-$30\Myr$, $30$-$100\Myr$, plus additional $5$ logarithmically spaced bins between $100\Myr$ and the age of the Universe), using the `continuity' prior described in \citet{Leja+19} to avoid abrupt discontinuities.
We adopted the dust attenuation model of \citet{CF00}, and include in the model spectra nebular emission lines by fixing the ionisation parameter to $\log(U)\!=\!-3$.
Templates for active galactic nuclei are not included in our models. To infer their luminosity distance, all galaxies are considered at the redshift of their host clusters, listed in Table \ref{table_data}. 
The resulting $M_\star$ include stellar remnants but exclude the gas lost by stellar winds and supernovae. SFR measurements are averaged over the latest $100\Myr$. 

\subsection{Stacking}
We employ a spectral stacking routine \citep{Sinigaglia2022c, Bianchetti2025a} to expand our analysis below the direct detection limit of the dataset. Specifically, we stack cluster galaxies with known optical spectroscopic redshift, regardless of their individual HI properties, and compare their mean HI content with those of the field population from the GALEX Arecibo SDSS Survey sample \citep[xGASS,][]{Catinella_2018}.

We briefly outline the stacking and procedure, but refer to \cite{Sinigaglia2022c} and \cite{Bianchetti2025a} for an exhaustive description. We start from the spectroscopic information provided by a catalog of cluster galaxies, inherited from the WINGS+OmegaWINGS dataset \citep{Moretti2014, Moretti2017}. We select only galaxies that are cluster members, defined to have a spectroscopic redshift within 3 times the cluster velocity dispersion from the mean cluster redshift.

For each galaxy, a three-dimensional HI cubelet was extracted from the parent datacube, centered on the optical position and spectroscopic redshift. The spatial extraction aperture spans a size of 100 kpc, ensuring that the full HI emission is captured regardless of source size or resolution. Along the spectral axis, the cubelet covers approximately a $\pm 2000$ km s$^{-1}$ range around the expected line position. Each cubelet was collapsed over the spatial dimensions to obtain a one-dimensional spectrum. Spectra are shifted to the rest frame using the corresponding spectroscopic redshift, and then rebinned onto a common velocity grid with fixed bin width to ensure homogeneous sampling and flux conservation. Flux densities were then converted into HI mass spectra using the standard relation from \citet[][]{Meyer_2017}. Once extracted, all spectra are primary beam corrected for the transmission factor $f$ and assigned a weight $f^2$. The final stacked spectrum was integrated over a fixed velocity window of $\pm 150$ km s$^{-1}$ to obtain the average HI mass. The integration range was defined as the narrowest interval encompassing the full width of the stacked line, while excluding the noise pattern, used to estimate the uncertainty. We also assigned a $1\sigma$ uncertainty to the HI mass estimate by computing the
RMS of the channels in the stacked spectrum, i.e. those that lie outside the integration range. The integrated signal-to-noise ratio, S/N, of the final stacked spectrum is computed as the ratio between the mass estimate and the $\sigma$ value defined above, assessing the statistical significance of the measurement. We consider a stacked line to be detected if it reaches S/N$>3$, following the criterion introduced by \citep{Sinigaglia2022c} and \citep{Bianchetti2025a}. In this analysis, all detections are characterized by a S/N$\geq5$.

 To ensure completeness and guarantee a meaningful statistical comparison with the comparison sample from xGASS, we applied a stellar mass cut ($9<\log_{10}{(M_{\star}/M_{\odot})}<11$) for the stacking analysis, that resulted in a sample of 677 cluster galaxies. As both stellar mass and cluster-centric distance are known to positively correlate with the HI mass in galaxies, we explore separately their effect by stacking galaxies in different $M_\star$ or $R/R_{200}$ bins. The cluster sample was then compared with xGASS field galaxies. The comparison with xGASS is drawn as follow. For each bin in the cluster sample, either in stellar mass or cluster-centric distance, we extracted a subsample of random galaxies from xGASS to match the bin stellar mass distribution, and measured the average HI mass of the realization. This process was repeated 10 times, every time with randomly extracted values, to derive the average HI mass between the 10 different realizations, $M_\text{HI}^\text{xGASS}$. We further defined a sub sample of star-forming (SF) galaxies, for both the cluster and the xGASS sample. For the cluster galaxies, star forming systems are defined on the basis of their emission lines, whereas for xGASS galaxies, due to the lack of spectroscopic information, we classify as `star-forming' those systems with a specific star formation rate higher than $10^{-11}$ yr$^{-1}$. The average HI mass content of each bin is reported in Table \ref{tab_stack}, whereas the individual stacked spectra are shown in Figure \ref{stack_spectra}. The integrated signal-to-noise of the stacked HI spectra ranges between 5 and 16.

\begin{table}[]
\caption{\label{tab_stack}Stacking analysis results.}
\setlength{\tabcolsep}{4pt} % Default value: 6pt
\renewcommand{\arraystretch}{1.2} % Default value: 1
  \centering
  \begin{tabular}{llccccc}
  \toprule
  \toprule
$x$&Sub.&S/N & N$_\text{gal}$ & $\widehat{x}$ & $M_\text{HI}$ & $M_\text{HI}^\text{xGASS}$\\
\midrule
\multirow{ 6}{*}{log$_{10}\left(\frac{M*}{M_\odot}\right)$}&\multirow{ 3}{*}{Full}&9.01&214&9.21&8.48$\pm$0.05&9.06$_{-0.03}^{+0.03}$\\
&&5.82&202&9.72&8.37$\pm$0.07&9.20$_{-0.03}^{+0.01}$\\
&&11.43&261&10.44&8.48$\pm$0.04&9.24$_{-0.06}^{+0.01}$\\
\cmidrule{2-7}
&\multirow{ 3}{*}{SF}&13.03&92&9.19&8.89$\pm$0.03&9.07$_{-0.02}^{+0.05}$\\
&&9.25&98&9.72&8.73$\pm$0.05&9.38$_{-0.01}^{+0.03}$\\
&&13.68&81&10.36&9.01$\pm$0.03&9.53$_{-0.07}^{+0.05}$\\
\midrule
\multirow{6}{*}{$\frac{R}{R_{200}}$}&\multirow{3}{*}{Full}&5.14&239&0.38&8.20$\pm$0.08&9.15$_{-0.02}^{+0.05}$\\
&&10.11&234&0.75&8.62$\pm$0.04&9.15$_{-0.05}^{+0.03}$\\
&&9.57&204&1.04&8.62$\pm$0.05&9.15$_{-0.03}^{+0.04}$\\
\cmidrule{2-7}
&\multirow{3}{*}{SF}&10.13&63&0.46&8.77$\pm$0.04&9.28$_{-0.06}^{+0.04}$\\
&&15.83&104&0.77&8.90$\pm$0.03&9.33$_{-0.05}^{+0.02}$\\
&&12.30&104&1.04&8.97$\pm$0.04&9.37$_{-0.02}^{+0.02}$\\ 
\bottomrule
  \end{tabular}
  \tablefoot{From left to right: Stacked quantity; Sub-sample; Stack signal-to-noise ratio; Number of galaxies per bin; median bin value; Mean HI mass; Mean HI mass in xGASS equivalent bin. HI masses are expressed in units of log$_{10}(M_\text{HI}/M_\odot)$. }
  \label{tab:placeholder}
\end{table}

\section{Results}
\label{section2}
\begin{figure*}[th!]
\sidecaption %
  \includegraphics[width=12cm]{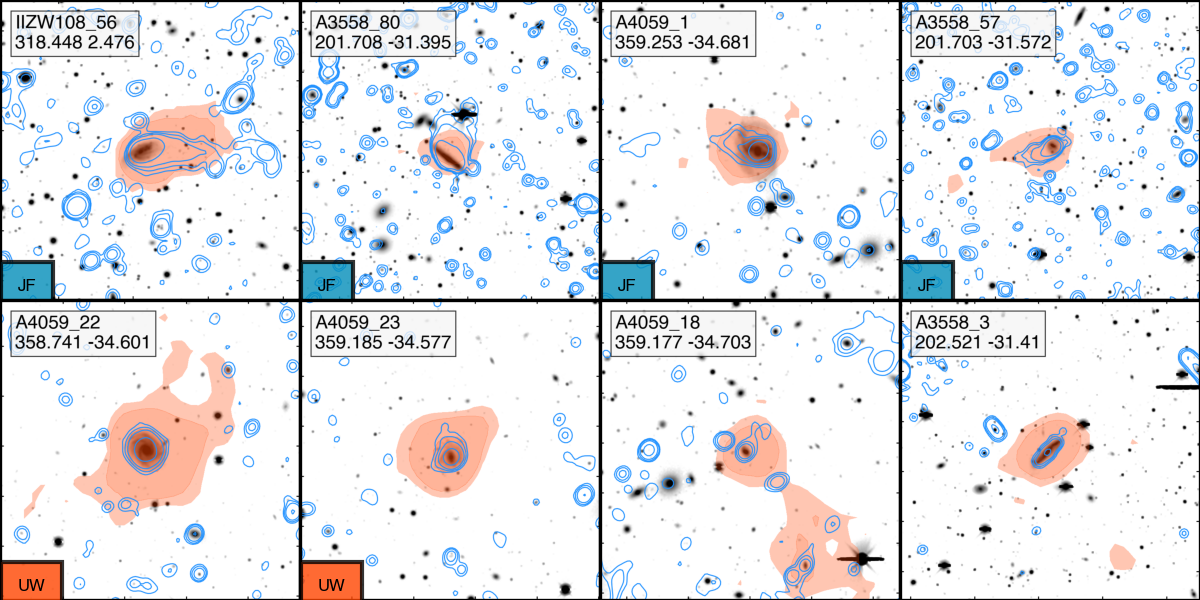}
  \caption{Examples of detected galaxies. For each galaxy we show a 6 arcmin cutout ($\sim360$ kpc at cluster redshift) of Legacy R-band images with the MeerKAT HI emission (red filled contours) and radio continuum (blue contours) at the 3, 6, 12, 24$\sigma$ levels on top. In the top-left corners are reported galaxy ID and optical center J2000 coordinates. Unwinding (UW) and jellyfish (JF) galaxies are labeled accordingly in the bottom-left corner.}
  \label{mosaic_1}
\end{figure*}
\begin{table}[]
\caption{\label{tab_class}Sample size per cluster and corresponding distribution in the analyzed categories.}
  \centering
  \begin{tabular}{cccccccc}
\toprule
\toprule
Cluster&N$_{\text{gal}}$ & First & Rec. & Int. & Late & JF & UW\\
\midrule
IIZW108&11&0&2&1&0&2&0\\
A4059&12&1&4&1&1&2&2\\
A3558&38&0&5&14&1&9&8\\
\midrule
Total & 61&1&11&16&2&13&10\\
\bottomrule

  \end{tabular}
  \tablefoot{From left to right: number of galaxies per cluster, inluding the three GASP galaxies; number of galaxies for each phase-space class: first, recent (Rec.), intermediate (Int.) and late ; number of jellyfish (JF) and unwinding candidates (UW). }
  
\end{table}
We first focus on our HI+radio continuum selected sample composed by galaxies (1) without close companions in either the optical or HI emission, which resulted in a blending of HI spectra, or without blending with back- or foreground sources in radio continuum, (2) having an HI emission and radio continuum observed with a significance above the 3$\sigma$ level and, for the radio continuum emission, an angular size larger than the corresponding angular resolution, and (3) with a reliable SED fit. We note that setting a lower limit on the radio continuum flux density resulted in a lower cut in radio continuum luminosity $L_\text{R}\gtrsim10^{27}$ erg s$^{-1}$, corresponding to SFR$\gtrsim0.1$ $M_\odot$ yr$^{-1}$. These criteria reduce the sample size from the original 116 HI-detected galaxies down to 61. The sample includes three well-studied galaxies from the GASP sample \citep[GAs Stripping Phenomena in galaxies with MUSE,][]{Poggianti2025} with extensive H$\alpha$ tails, namely JO206 in IIZW108 (IIZW108$\_$56), JO194 in A4059 (A4059$\_1$), and JO147 in A3558 (A3558$\_$80), and four other galaxies from the GASP sample with less extended extraplanar emission in the MUSE data, namely JO144 (A3558$\_103$), JO153 (A3558$\_24$), JO159 (A3558$\_41$) and JO205 (IIZW108$\_7$). Although the low angular resolution does not permit us to investigate the individual morphological features of each detected galaxy, as a notable feature we report the discovery of a $\sim100$ kpc HI tail extending from the galaxy  A4059$\_$18. In Figure \ref{mosaic_1} we show a selection of galaxies from the sample, whereas the full sample is presented in Figure \ref{mosaic}. In those figures, the HI (red-filled contours) and radio continuum (blue contours) emissions are super-imposed on Legacy R-band images.

\begin{figure}
  \centering
  \includegraphics[width=\linewidth]{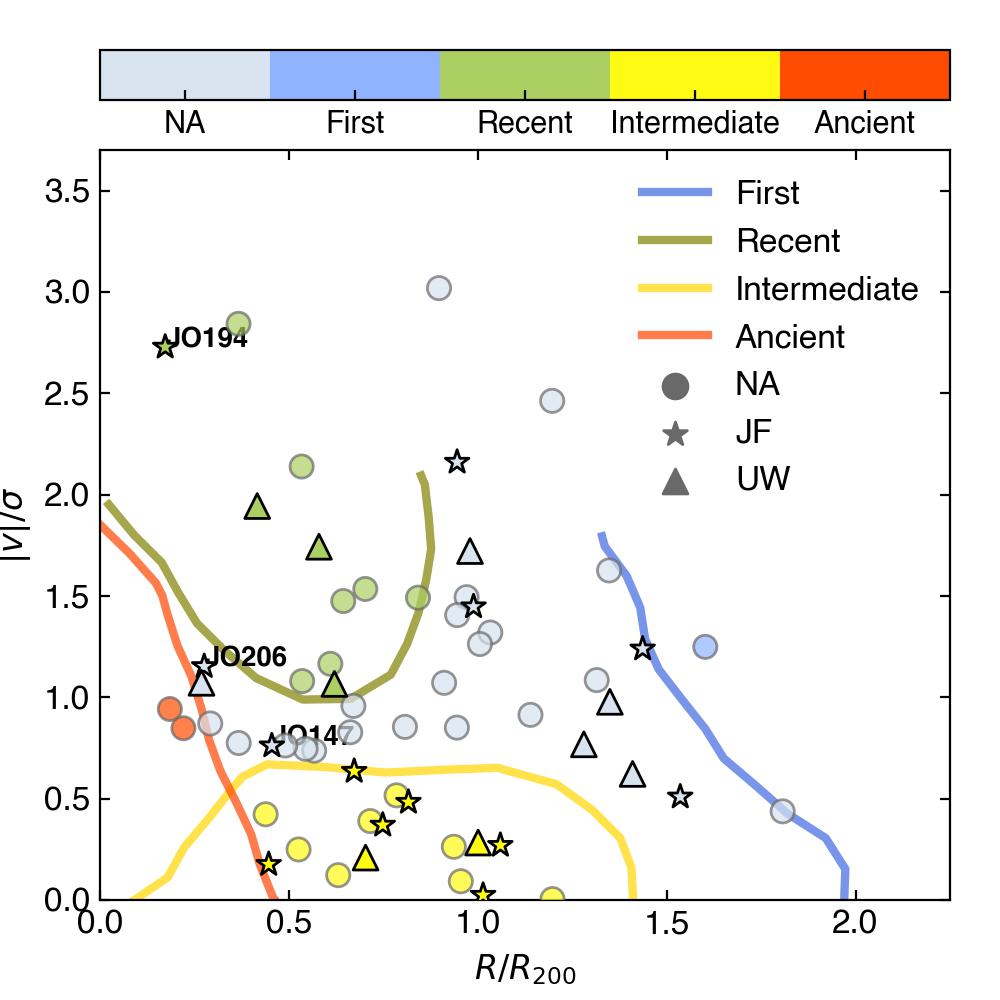}
  \caption{Phase-space diagram. Galaxies are color-coded for their membership in the different infalling stage categories. UW and JF galaxies are represented with, respectively, triangles and stars. Colored lines mark the limits of the different categories taken from \citet{Rhee_2017}.}
  \label{phase-space}
\end{figure}

The galaxies have been further classified by using two criteria, i.e. the presence of specific features in their optical morphology such as the presence of extraplanar debris or irregular spiral arms pattern, and the position in the phase-space diagram given by their projected line-of-sight velocity, measured from the HI spectrum, and cluster-centric distance (Figure \ref{phase-space}). The first classification comes from the cross-matching with the jellyfish galaxy candidates \citep[JF,][]{Poggianti2016} and unwinding catalogs \citep[UW,][]{Vulcani_2022}. The second criterion makes use of the phase-space classification presented by \citet[][]{Rhee_2017}. Based on cosmological simulations, the projected clustercentric distance and line-of-sight velocity of a galaxy can be used to statistically assign the infalling stage between first infallers (blue), which have not yet entered the cluster, recent (green), which have fallen into the cluster in the past 3.6 Gyr, intermediate (yellow), which fell into the cluster between 3.6 and 6.5 Gyr ago, and ancient infallers (red), that have been in the cluster for more than 6.5 Gyr. Galaxies which could not be assigned to any classes, either for the morphological classification or the phase-space one, are labeled as NA. The number of galaxies for each class are reported in Table \ref{tab_class}, whereas their individual properties are reported in Table \ref{long_tab}. 

\begin{figure*}
  \centering
  \includegraphics[width=1\linewidth]{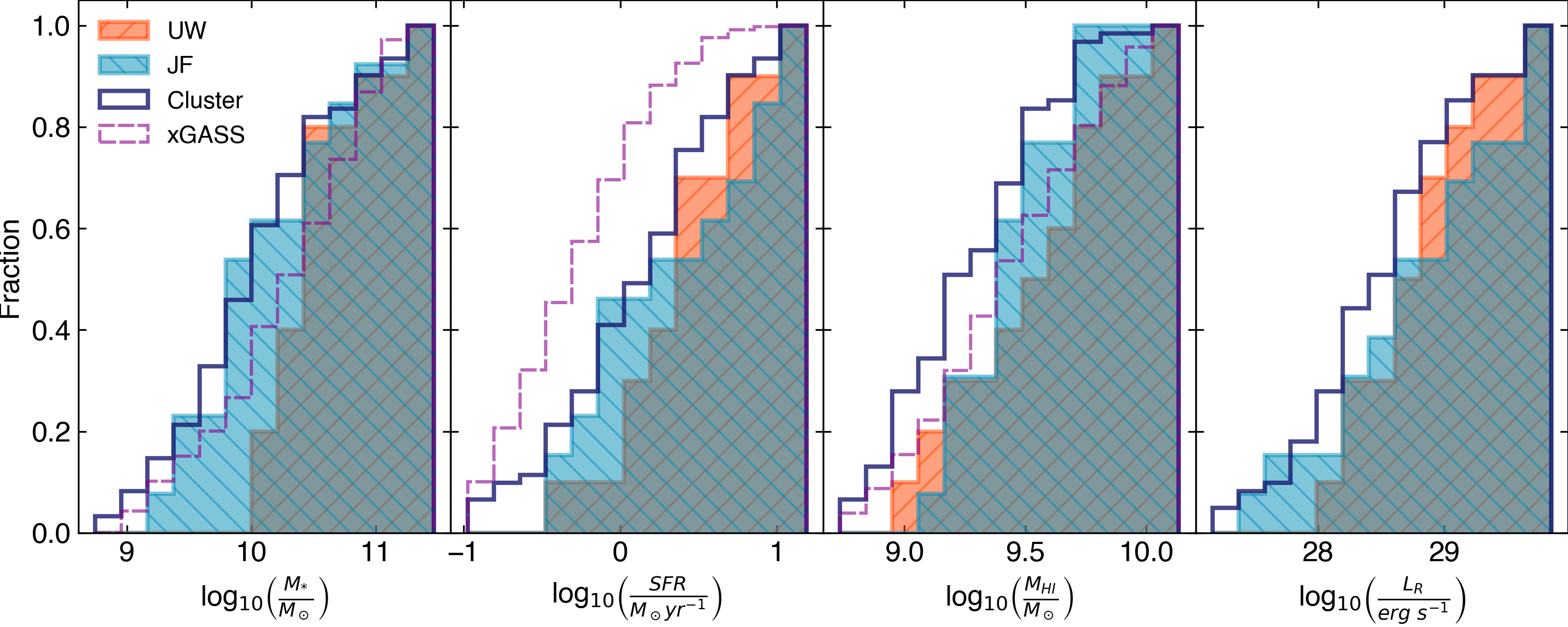}
  \caption{From left to right: cumulative distribution of measured stellar mass, star formation rate, HI mass, and radio continuum luminosity for the cluster sample (dark-blue line), UW (orange-filled), JF (teal-filled) and the field sample drawn from xGASS (purple dashed line). }
  \label{dist}
\end{figure*}

The cumulative distributions of stellar mass, SFR, HI mass and radio continuum luminosity are shown in Figure \ref{dist}. For comparison, we also show the cumulative distribution of an xGASS subsample derived by applying a lower cut in HI mass and SFR to match our sample's HI mass and radio continuum luminosity detection limits. xGASS SFR values have been corrected to match our assumed IMF. This comparison reveals that within similar $M_{\text{\ion{H}{i}}}$ ranges, the cluster galaxies are $\sim2\times$ less massive compared to the field sample. However, cluster galaxies are, on average, a factor $\sim2.7$ more star-forming than the field ones. To confirm this difference we perform the Kolmogorov-Smirnov test between cluster and field stellar mass, HI mass and SFR distributions, finding p-values of, respectively, $2\times10^{-3}$, $4\times10^{-3}$, and $2\times10^{-5}$, that would reject the hypothesis that the two samples were drawn from the same distribution. This result is further discussed in Section \ref{meerkat_results}.

\begin{figure}
  \centering
  \includegraphics[width=\linewidth]{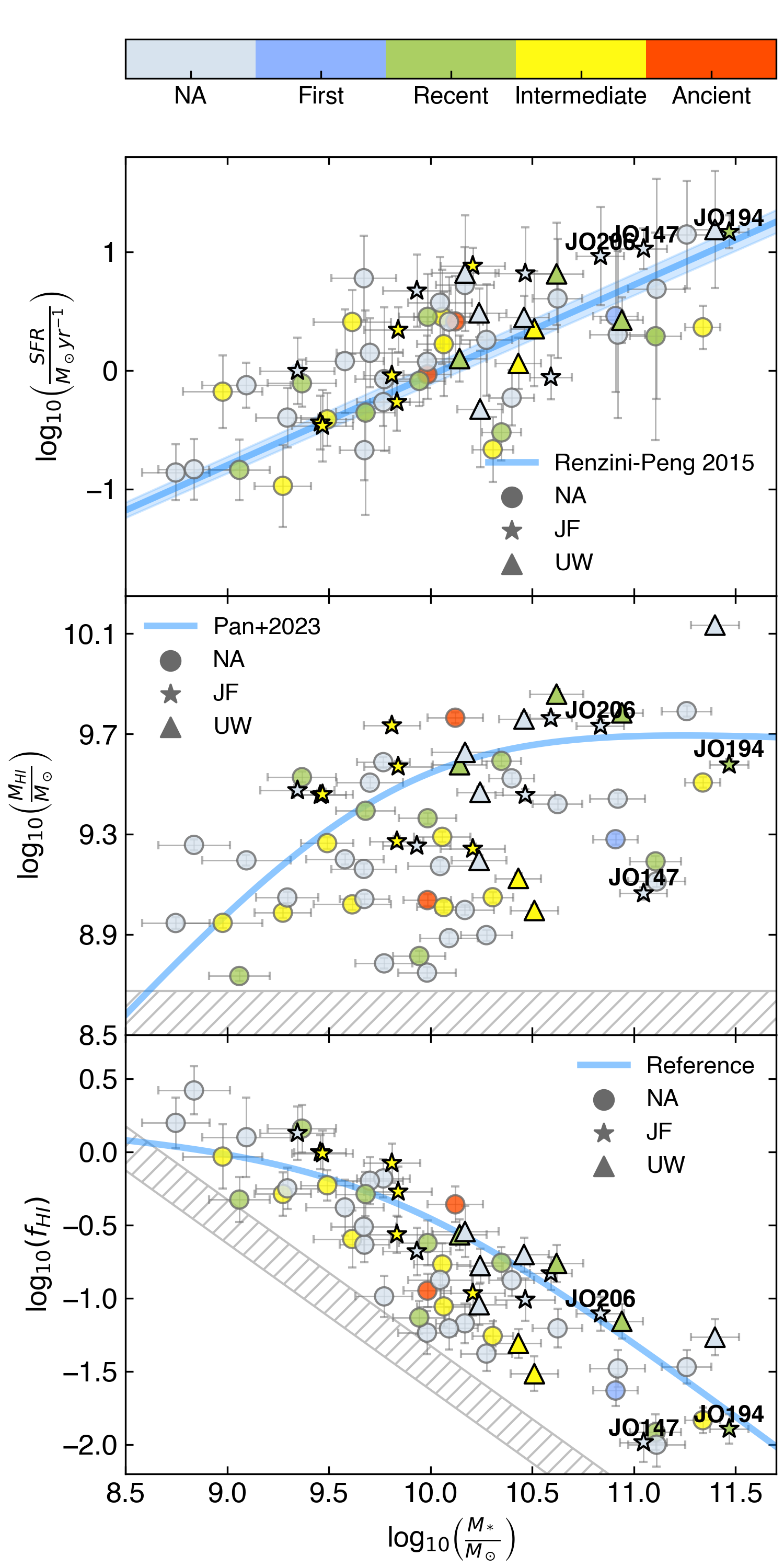}
  \caption{Star formation rate (top), HI mass (center) and HI mass fraction (bottom) vs. stellar mass. Galaxies are color-coded for their membership in the different infalling stage categories. UW and JF galaxies are represented with, respectively, triangles and stars. The blue lines represent the reference relations with the corresponding uncertainties (see text). In the central and bottom panel, the gray dashed area indicates the detection limit. }
  \label{paper1}
\end{figure}

\begin{figure}
  \centering
  \includegraphics[width=\linewidth]{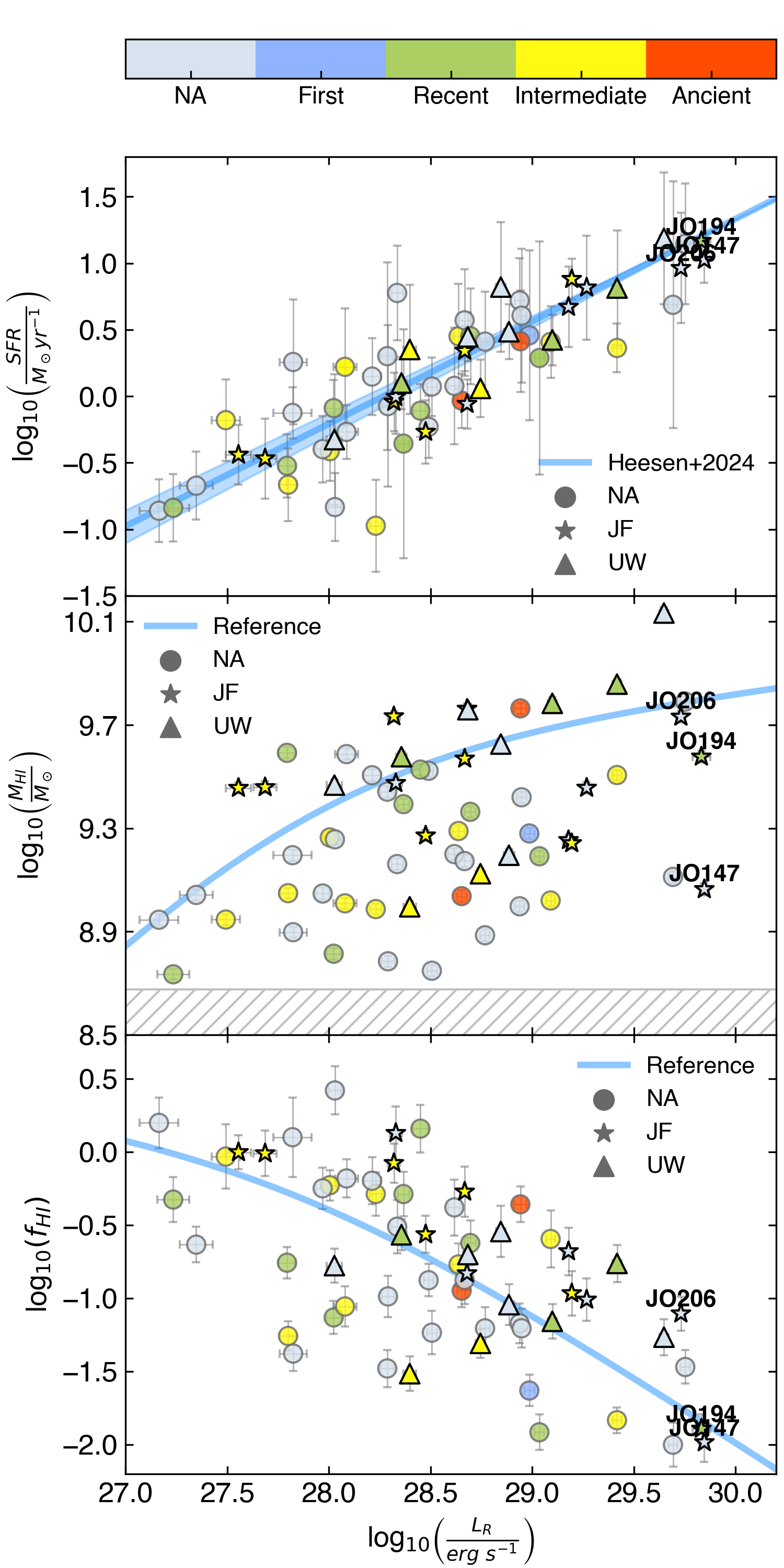}
  \caption{Star formation rate (top), HI mass (center) and HI mass fraction (bottom) vs. radio continuum luminosity. Galaxies are color-coded for their membership in the different infalling stage categories. UW and JF galaxies are represented with, respectively, triangles and stars. The blue lines represent the reference relations with the corresponding uncertainties (see text). In the central, the gray dashed area indicates the detection limit. }
  \label{paper2}
\end{figure}

In Figure \ref{paper1} and \ref{paper2}, the cluster galaxies are compared with reference relations derived on field galaxies, namely the SFR-$M_*$ \citep[][]{Renzini_2015}, the $M_\text{HI}$-$M_*$, and corresponding $f_\text{HI}$-$M_*$ \citep[][]{Pan_2023}, where $f_\text{HI}=M_\text{HI}/M_*$ is the HI mass fraction, and the SFR-$L_\text{R}$ \citep[][]{Heesen_2024}. Stellar mass, SFR and HI mass are expected to jointly evolve in the secular evolution of galaxies, with the HI gas being converted into stellar mass. In this context, radio continuum luminosity, which is related to the galaxy nonthermal energy density in magnetic field and cosmic ray electrons, can be a proxy of the most recent star formation rate \citep[$\sim10^7$ yr, see][]{Condon_1992,Kennicutt_2012,Werhahn2021,Pfrommer2022}, or tracer for recent environmental processing \citep[][]{Ignesti_2022b, Edler_2024}. Galaxies are colored for their phase-space classification, whereas the symbols represent the optical classification. 

These reference relations can be combined to obtain derivative relations, such as the $f_\text{HI}-M_*$, $f_\text{HI}-L_\text{R}$, and the $M_\text{HI}-L_\text{R}$ relations which we show in Figure \ref{paper1} and \ref{paper2}. We observe that the galaxies broadly follow the SFR-$M_*$ relation, although the majority of galaxies lies above it. The JFs tend to lie slightly above the relation, which is consistent with the 0.2 dex SFR excess found by \citet{Vulcani_2018c}. We present a detailed discussion of UW and JF systems in Section \ref{uw-jf}. Although the SFR is normal or increased relative to their stellar mass, most of the sample lies below the HI-stellar mass relation. This effect is particularly visible for galaxies with $M_*>10^{10}~M_\odot$, because the HI masses expected for this $M_*$ range are well above the MeerKAT detection limit, which allows us to probe the "HI-deficient" regime more effectively. Therefore, only high mass galaxies can be detected in a stage of advanced HI deficiency. The new results for JO206 and the galaxies in A4059 are consistent with the previous corresponding VLA studies   \citep{Ramatsoku_2020,Luber_2022}. 

Regarding the radio luminosity, galaxies overall follow the expected relation with the SFR, albeit with a scatter of $\sim$0.3dex, largely driven by the uncertainties on the SFR, that are typically of 0.3dex. At any given radio luminosity, which is a proxy of SFR, we observe HI deficiency, indicating that $L_\text{R}$ is more closely linked to the SFR than the SFR is to the HI mass. Concerning $f_\text{HI}$, we observe a larger scatter in comparison with $L_\text{R}$ than in $M_*$, with the majority of points laying above the expected trend. Concerning the phase-space classification, we observe that relatively more intermediate systems than recent ones lie below the $M_\text{HI}-M_*$ relation (10/16 vs 4/11). We further note that the two galaxies classified as ancient infallers, hence that would have been subject to the cluster effects for the longest period, show a limited scatter with respect to the reference relations,  but we remind the reader that the observed radii and velocities in the phase-space diagram are both lower limits to the real, 3D values, hence they could be false-positive in the phase space classification due to projection effects.


\begin{figure}
  \centering
  \includegraphics[width=\linewidth]{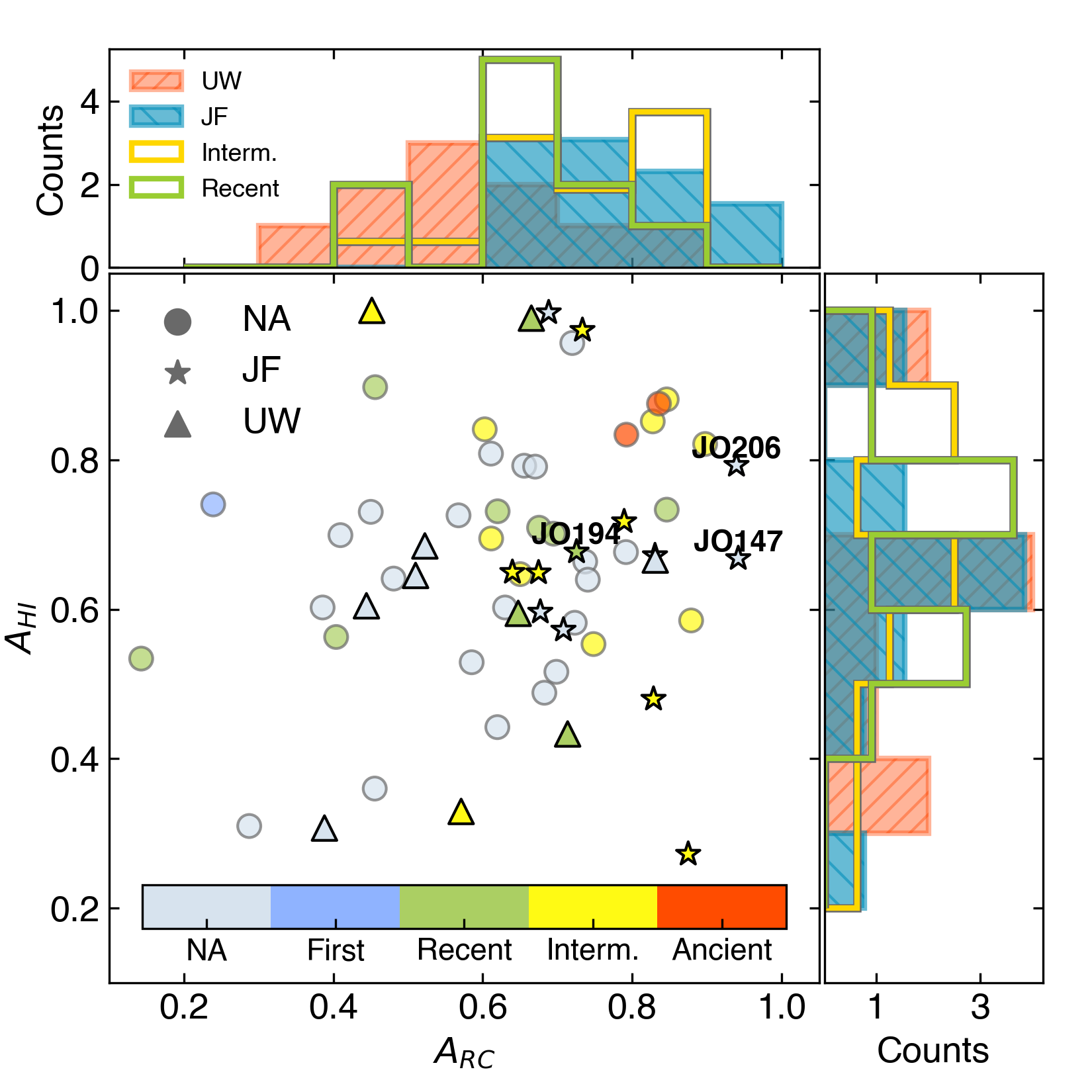}
  \caption{HI vs. radio continuum asymmetry. Along each axis are shown the corresponding distributions of the UW (orange-filled), JF (teal-filled), recent (green line) and intermediate (yellow line) sub- samples. Galaxies are color-coded for their membership in the different infalling stage categories. UW and JF galaxies are represented with, respectively, triangles and stars. }
  \label{A-A}
\end{figure}

In Figure \ref{A-A} we investigate the asymmetry parameters computed on the HI and radio continuum emission. The majority of galaxies appears to be highly disturbed, showing high values in both $A_\text{HI}$ and $A_\text{RC}$, that suggests that their ISM is being perturbed by the environment. As a caveat, we stress that the direct comparison between the two asymmetries is limited by the fact that 1) radio and HI images have different angular resolution, although they share the same beam size-to-pixel area ratio, and 2) each asymmetry is computed with respect to the corresponding barycenter. We adopted this approach because the projected physical distance between the optical, the radio continuum and HI centers are typically smaller than 10 kpc, which is smaller than the resolved physical scale of our images. This entails that the radio-optical shifts, which would strongly affect the measured asymmetry, cannot be reliably measured, nor disentangled from uncertainties related to the astrometry correction or local phase calibration. In Table \ref{A_res} we list the mean and standard deviation values of the HI and radio continuum asymmetry for the largest sub-samples, i.e. recent and intermediate infallers, UW and JF galaxies. 
\begin{table}[]
  \centering
  \caption{HI and radio continuum asymmetry}
  \begin{tabular}{ccc}
  \toprule
  Sample&$A_\text{HI}$&$A_\text{RC}$\\
  \midrule
  Full&0.67$\pm$0.18&0.64$\pm$0.17\\
  Recent&0.69$\pm$0.15&0.60$\pm$0.19\\
  Inter.&0.68$\pm$0.20&0.73$\pm$0.13\\
  UW&0.63$\pm$0.22&0.57$\pm$0.13\\
  JF&0.67$\pm$0.18&0.77$\pm$0.10\\
\midrule
  \end{tabular}
  \tablefoot{\label{A_res} Mean values and standard deviation of $A_\text{HI}$ and $A_\text{RC}$ for, from top to bottom, full sample, recent and intermediate infallers, UW and JF galaxies.}
\end{table}

\begin{figure*}
\sidecaption %\centering
 \includegraphics[width=12cm]{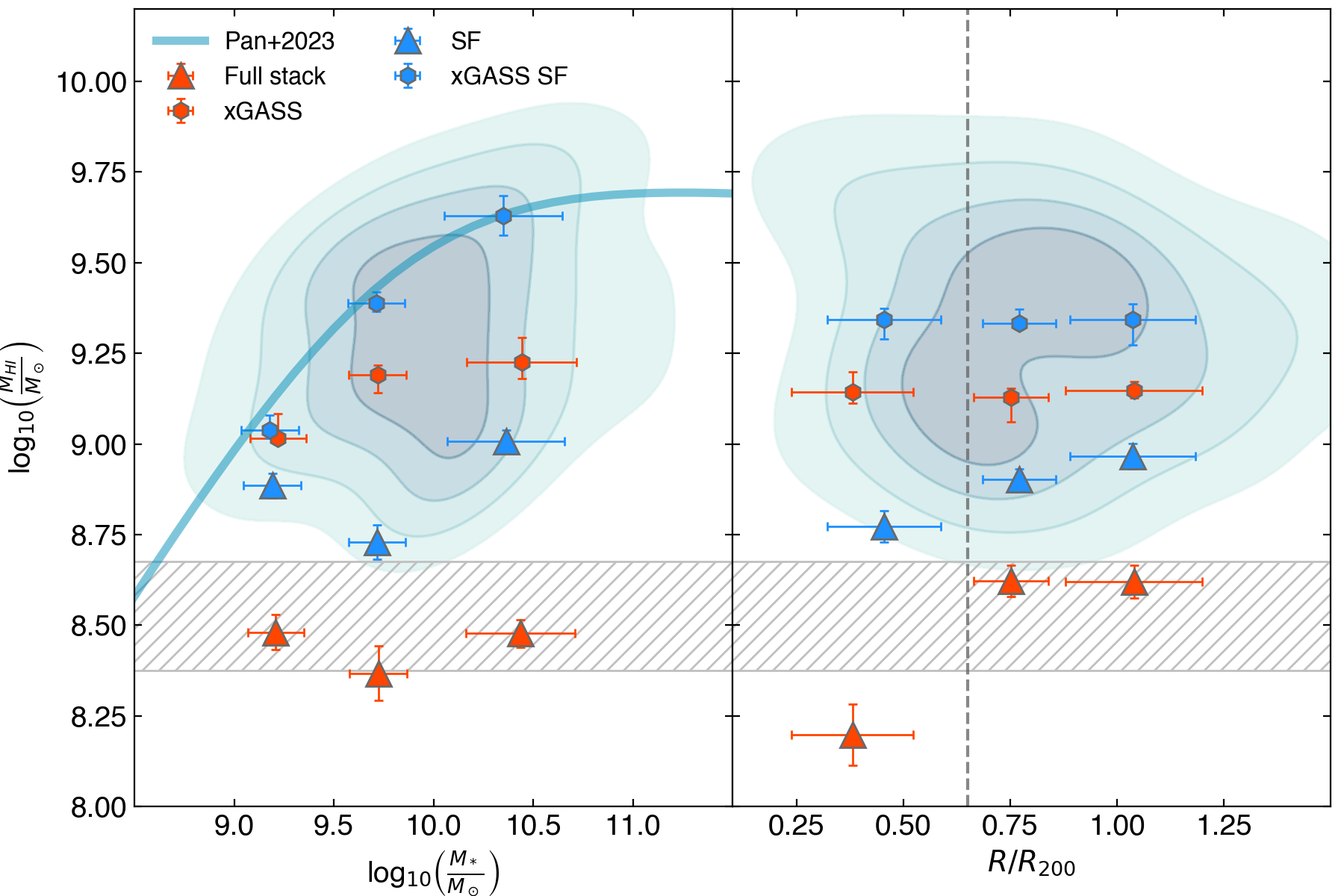}
  \caption{\label{stack}Stacked HI mass for bin in stellar mass (left) and cluster-centric distance (right) for cluster galaxies (triangles) for both the full (red) and star-forming (SF, blue) samples. For each cluster galaxy bin, hexagons indicate the average HI mass of xGASS galaxies sub-samples with matching stellar mass distributions. The blue line indicate the reference HI-stellar mass relation and the gray-dashed areas indicates the mass detection limit. Light-blue filled contours indicate the 2D probability density contours of the detected sample. The vertical dashed line in the right panel indicates $R_{500}$.}
\end{figure*}

Finally, in Figure \ref{stack} we show the results of the stacking analysis. The average HI mass of the stacked samples, binned for stellar mass (left panel) and cluster centric distance (right panel), are compared with the detected sample and the corresponding equivalent in stellar mass distribution from the xGASS sample. We observe that cluster galaxies have systematically less HI mass than the field sample at any stellar mass, and we find a decreasing trend with the cluster-centric distance. These results are further discussed in Section \ref{stack-disc}.

\section{Discussion}
\label{section3}
\subsection{The HI+radio continuum MeerKAT view}
\label{meerkat_results}

Cluster galaxies detected in both HI and radio continuum differ from field galaxies, whether comparing to xGASS or to field empirical relation, as they appear more actively star forming despite hosting less HI gas. They are also highly asymmetrical in both HI and radio continuum, indicating that the ram pressure is perturbing their ISM. This suggests that the combined HI and radio continuum selection produced a sample biased toward galaxies in a specific stage of their evolution within the cluster. As the cluster environment is expected to consume the HI content, these galaxies are likely in an initial RPS stage: they have lost only part of their HI and begun quenching, and the initial ICM compression on the ISM resulted in a temporary increase in SFR \citep[][]{Vulcani_2018c, Molnar_2022, Roberts_2022e}. Moreover, a part of the missing HI should have been converted into H$_2$ to fuel the star formation on time-scales typically shorter than observed in field galaxies \citep[][]{Moretti_2020b}.

We further observe that, at fixed stellar mass, intermediate infallers have lower \ion{H}{i} masses than recent infallers (32$\%$ and 62$\%$, respectively, are below the HI-stellar mass relation), that would be in line with the fact that the longer galaxies stay in the cluster, the more they lose gas due to the environment. We also note that, whereas $A_\text{HI}$ stays constant between recent and intermediate infallers, $A_\text{RC}$ shows a tentative increment (Table \ref{A_res} and Figure \ref{A-A}). This result suggests that galaxies become HI-asymmetric within the first few Gyr after they entered the clusters, whereas the radio continuum asymmetry takes longer to develop than the HI one. The large $A_\text{HI}$ asymmetry in the initial ram-pressure stripping process could be due to thermal instability of the stripped CGM \citep[][]{Sparre2024}. Due to the lack of relativistic electrons, the stripped CGM has no radio continuum emission, hence resulting in low $A_\text{RC}$ values. At later times, when the ISM is also efficiently stripped, galaxies develop also radio continuum tails, thus increasing $A_\text{RC}$. However, a larger statistic is required to confirm this hypothesis.

At 1.4 GHz, the radio continuum emission traces particles with a cooling time of the order of a few $10^7$ yr in the typical ISM magnetic intensity of $\sim10~\mu$G \citep[][]{Basu_2015, Chiu2025}, hence it can be used to disentangle the different stages of the early environmental processing. As previously discussed in the literature \citep[e.g.,][]{Ignesti_2022b,ArangoToro_2023}, the ratio between the observed and expected SFR-powered radio continuum luminosity, $L_\text{R}/L_\text{R}^\text{SFR}$ can depend on the recent star formation history. As the radio emission mainly depend on the SFR, here $L_\text{R}/L_\text{R}^\text{SFR}$ roughly corresponds to the ratio between the SFR in the last $10^7$ yr \citep[][]{Kennicutt_2012} and the average star formation over $\sim10^8$ yr. Therefore, if the SFR has been steadily decreasing over the past hundred Myr, the galaxy would show $L_\text{R}/L_\text{R}^\text{SFR}<1$. Conversely, a galaxy in which the SFR is surging in the past tens of Myr would show $L_\text{R}/L_\text{R}^\text{SFR}>1$. However, a similar luminosity ratio can also be due to the fact that the galaxy is more radio luminous than expected due to the radio continuum emission produced by old relativistic electron still present in the galaxy \citep[][]{Ignesti_2021,Ignesti_2022b,Edler_2024}. 

\begin{figure}
    \centering
    \includegraphics[width=\linewidth]{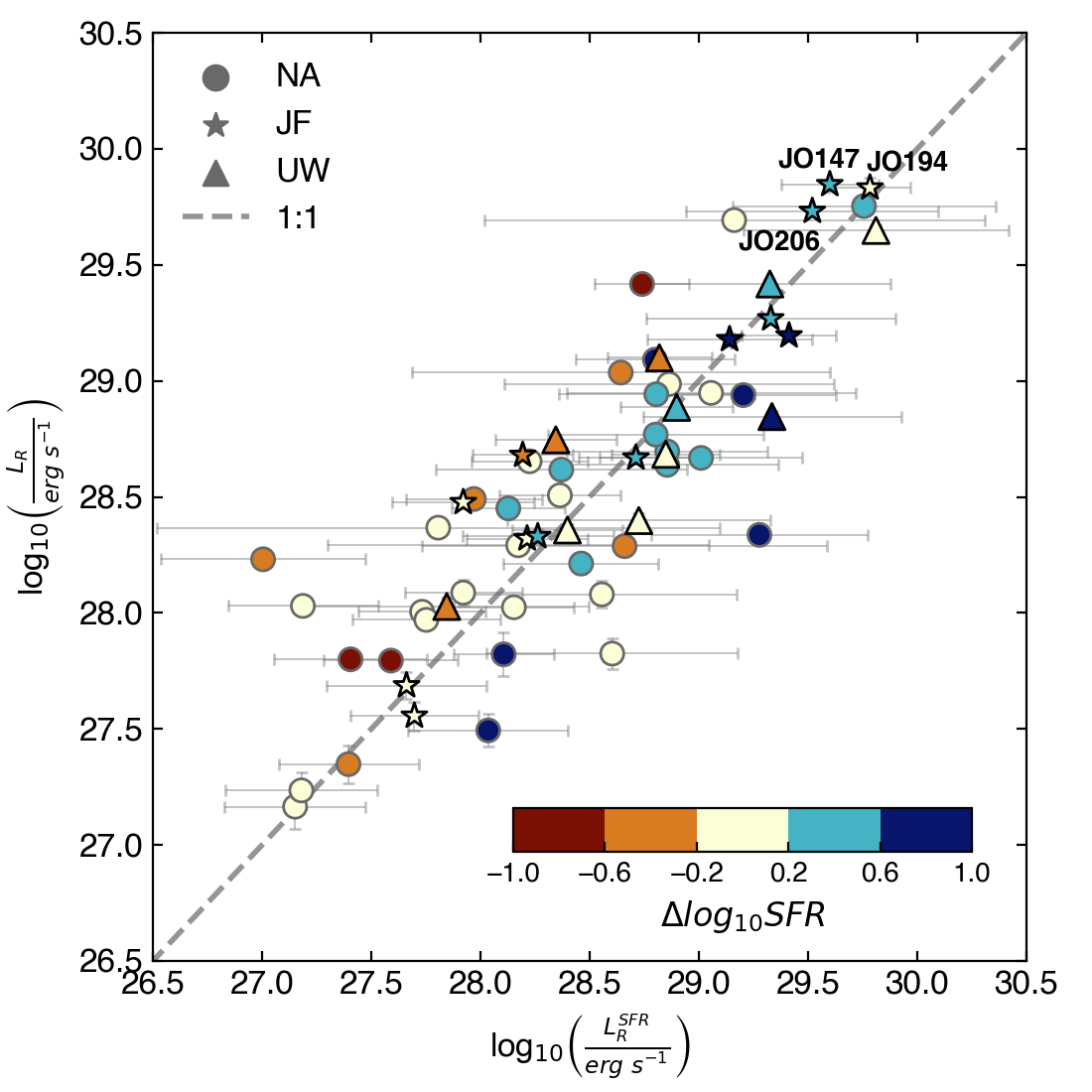}
    \caption{Observed vs. expected radio continuum luminosity. Galaxies are color-coded for $\Delta\text{log}_{10}\text{SFR}$. UW and JF galaxies are represented with, respectively, triangles and stars. The dashed line indicates the 1:1 relation.}
    \label{paper4}
\end{figure}

In Figure \ref{paper4} we compare the observed radio luminosity with the one expected from the SFR, $L_\text{R}^\text{SFR}$, color-coded for their distance from the \citet[][]{Renzini_2015} star formation sequence, $\Delta\text{log}_{10}\text{SFR}$ (Figure \ref{paper1}, top panel). The two luminosities show a good correlation, as expected in star forming galaxies, but it emerges that the scatter is driven by $\Delta\text{log}_{10}\text{SFR}$. Specifically, galaxies with $\Delta\text{log}_{10}\text{SFR}>0$ show $L_\text{R}^\text{SFR}\geq L_\text{R}$, whereas galaxies more quenched ($\Delta\text{log}_{10}\text{SFR}<0$) preferentially show $L_\text{R}^\text{SFR}< L_\text{R}$. 

\begin{figure}
  \centering
  \includegraphics[width=\linewidth]{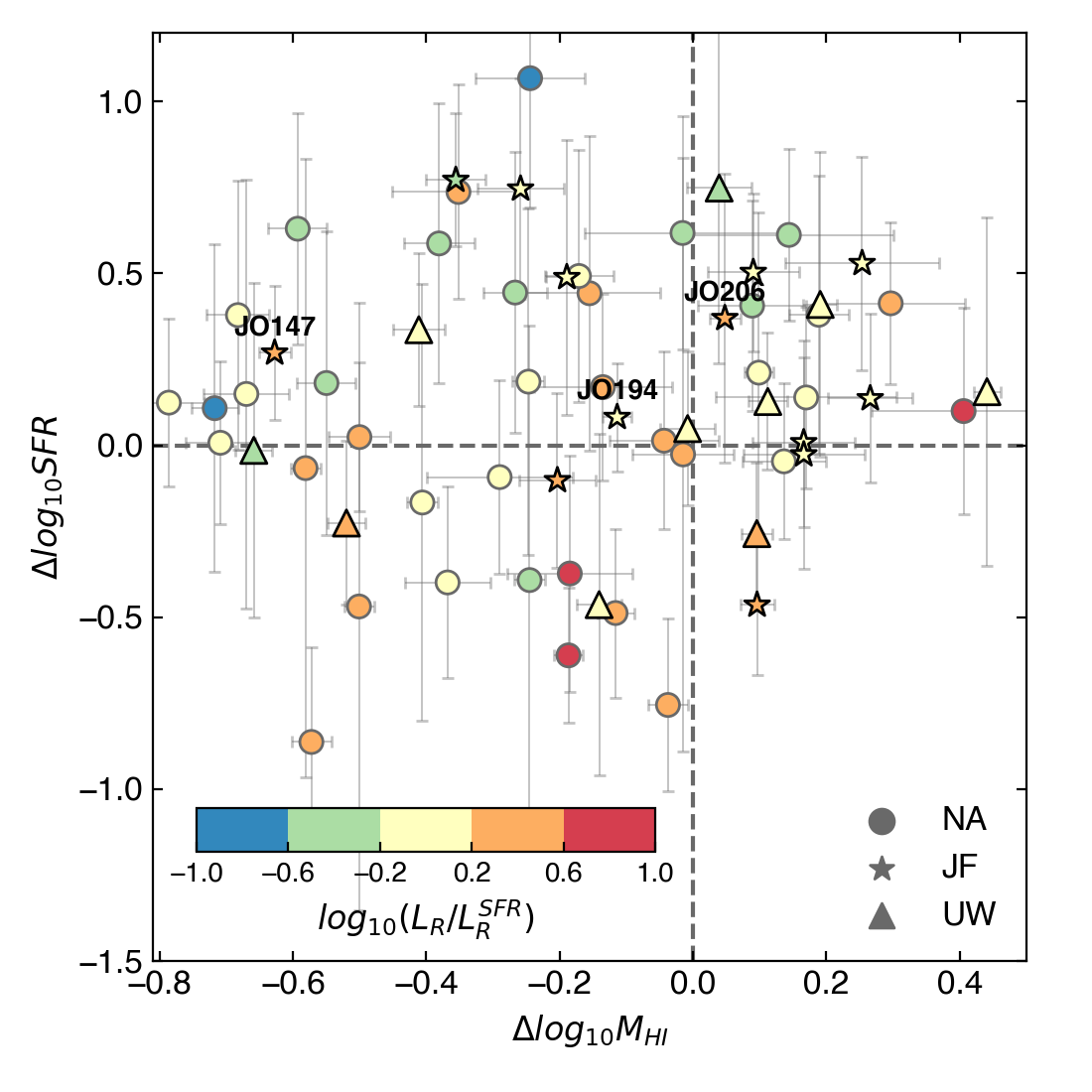}
  \caption{Logarithmic differences between the observed SFR and $M_\text{HI}$ and the reference relations presented in the text, color-coded for $\text{log}(L_\text{R}/L_\text{R}^\text{SFR})$.  UW and JF galaxies are represented with, respectively, triangles and stars. For reference, dashed lines indicate the zeroth-levels.}
  \label{def}
\end{figure}

We then argue that $L_\text{R}/L_\text{R}^\text{SFR}$ and $\Delta\text{log}_{10}\text{SFR}$ can be used to define the subsequent stages of the early processing:
\begin{enumerate}
    \item $L_\text{R}/L_\text{R}^\text{SFR}\simeq1$ and $\Delta\text{log}_{10}\text{SFR}\simeq0$: galaxies not yet affected by the environment;
    \item $L_\text{R}/L_\text{R}^\text{SFR}>1$ and $\Delta\text{log}_{10}\text{SFR}>0$: galaxies which are now undergoing the SFR burst;
    \item $L_\text{R}/L_\text{R}^\text{SFR}<1$ and $\Delta\text{log}_{10}\text{SFR}>0$: galaxies which have passed the compression-induced SFR burst for more than a few $\sim10^7$ yr. The SFR is still above the main sequence but the radio emission is no longer affected by the past burst;
    \item $L_\text{R}/L_\text{R}^\text{SFR}>1$ and $\Delta\text{log}_{10}\text{SFR}<0$: galaxies which have entered in their quenching phase, where their radio emission is possibly sustained by the old electron populations;
    \item $L_\text{R}/L_\text{R}^\text{SFR}<1$ and $\Delta\text{log}_{10}\text{SFR}<0$: Quenched systems with faint residual radio emission.
\end{enumerate}
To test this scenario, in Figure \ref{def} we compare $\Delta\text{log}_{10}\text{SFR}$ with difference between the observed and expected $M_\text{HI}$, $\Delta\text{log}_{10}M_\text{HI}$. \citet[][]{Molnar_2022} showed that cluster galaxies in the Coma cluster, due to the continuous gas loss, follow specific evolutionary tracks, moving from an initial HI mass-SFR sequence to a stage in which they show be HI-deficit associated with high SFR (top-left quadrant) to finally end in both HI and SFR deficit (bottom-left quadrant). First, we note that 23 galaxies, corresponding to $38\%$ of the sample, are simultaneously HI-poor but with a SFR above the main sequence (top-left quadrant in the left panel in Figure \ref{def}). These results are in line with the previous comparison with the xGASS sample, and they further indicate that the cluster influence can indeed result in a temporary increase in star formation even if the galaxy is loosing its HI gas, which is at odd with the typical evolution of field galaxies. 

Furthermore, we observe that galaxies with $L_\text{R}/L_\text{R}^\text{SFR}>1$ are preferentially HI-deficient ($\Delta\text{log}_{10}M_\text{HI}<0$), because they are either going through the ISM compression ($\Delta\text{log}_{10}\text{SFR}>0$) or they have begun quenching ($\Delta\text{log}_{10}\text{SFR}<0$) due to the gas loss. Similarly, we detect galaxies in which the SFR, albeit still above the expected values, is now declining ($L_\text{R}/L_\text{R}^\text{SFR}<1$) due to the gas loss ($\Delta\text{log}_{10}M_\text{HI}<0$). The emerging picture, in line with the previous studies, is that galaxies would move from the upper-right quadrant to the upper-left one and, eventually, end in the lower-left one. During this transformation, the ratio between radio luminosity and SFR changes reflecting the internal evolution. As previously pointed out in \citet{Ignesti_2021, Ignesti_2022b,Edler_2024}, we note that the $L_\text{R}/L_\text{R}^\text{SFR}$ ratio depends on both the observed frequency and the SFR tracer, which can both mitigate or enhance the discrepancy. We further observe that, to a small extent, the radio excess could be influenced also by low-luminosity AGN activity \citep[e.g.,][]{Panessa_2019,Peluso_2026}.

Finally, we note that the luminosity ratio can also be influenced by environmental effects, such as shock or compression, which should enhance the galaxy radio luminosity \citep[][]{Vollmer_2004,Murphy2009,Chen_2020}, or enhanced ionization losses along the galaxy leading side \citep[][]{Ignesti_2021}, that can reduce the low-energy electrons radio luminosity. Different processes would produce characteristic signatures in the radio continuum spectrum: if the radio excess is caused by shock relativistic electrons re-acceleration or subsonic ISM magnetic field compression, then galaxies would show a flat spectral index, otherwise, if the radio emission is maintained by old electron populations, then galaxies would show a steep spectral index, as previously shown by \citet[][]{Edler_2024}. In our case, additional observations at lower frequencies would be required to test the presence of such spectral index signatures.

\subsection{Unwinding vs Jellyfish galaxies: different stages of the same evolutionary process?}
\label{uw-jf}
In the previous section we have shown that the HI+radio continuum selection collected galaxies in the initial environmental processing stages. UW and JF galaxies are peculiar systems whose properties result from interacting with the environment, and previous works suggested that indeed the unwinding phase happens at the begin of the galaxy infall and it precedes the formation of the extraplanar tail \citep[][]{Bellhouse_2021,Vulcani_2022,Lassen2026b}. 

Our data partially support this evolutionary sequence. The phase-space analysis indicates that the ratio between intermediate and recent galaxies is 6:1 for the JF and 2:3 for the UW, supporting the scenario in which they are successive stages of the environmental processing. 
In terms of phase-space classification, UW galaxies show a ratio between intermediate and recent of 2:3, whereas for JF it is 6:1. This increment suggests that galaxies first evolve into UW in the first few Gyrs, and then they become JF. Correspondingly, UW and JF galaxies show similar mean $A_\text{HI}$ values of, respectively, $0.63\pm0.22$ vs $0.67\pm0.18$, but tentatively different $A_\text{RC}$, with JF showing, on average, $A_\text{RC}=0.77\pm0.10$ whereas UW systems show $A_\text{RC}=0.57\pm0.13$. As discussed previously, the difference in $A_\text{RC}$ could be due to the fact that JF galaxies have stayed in the cluster for longer than the UW galaxies. As a caveat, we note that JF are more easily identified in edge-on galaxies, whereas galaxies classified as UW are preferentially observed in face-on orientations, as it facilitates the characterization of their spiral arms and corresponding pitch angles. This entails that extraplanar asymmetries would be intrinsically more evident in JF than in UW systems. Moreover, a galaxy that is simultaneously UW and JF may be assigned one class or the other depending on the view angle and RPS angle. Indeed, in the GASP sample, there are indications that a number of extreme JF galaxies are also unwinding \citep[][]{Bellhouse_2021,Poggianti2025}. 

However, Figure \ref{dist} also shows that UW galaxies are massive galaxies with $M_*>10^{10}~M_\odot$ and, on average, the highest HI mass in the sample, whereas JF systems show a broader range in stellar mass and HI content. At first glance, the differences in HI and stellar mass between UW and JF galaxies may suggest that the duration of the unwinding phase \citep[$\sim0.5-0.8$ Gyr, see][]{Bellhouse_2021,Machado_2025} depends on the stellar mass, such that only massive galaxies remain in this phase long enough to be detected. We observe similar fraction of JF and UW galaxies, thus this scenario would imply that the UW phase duration drops dramatically for galaxies with $M_*<10^{10}~M_\odot$, possibly because their own gravitational potentials are too shallow to sustain the spiral structure against the environmental perturbation. Alternatively, given that a high fraction of UW galaxies seem to show stellar bars \citep[][]{Lassen2026b}, we suggest that part of the UW galaxies may instead represents a distinct evolutionary pathway for massive, baryon-dominated systems that are perturbed by external processes, instead of the intermediate evolutionary stage before becoming JF systems. 

A particularly extreme manifestation of the effects described above is observed in the three GASP jellyfish galaxies. These systems are SFR- and radio-enhanced, but HI-poor. While they all exhibit extraplanar radio continuum emission, only JO206 and JO194 also show extraplanar HI emission.
More generally, both the three GASP galaxies and JF systems show high radio continuum luminosity (Figure \ref{dist}, \ref{paper4}). We suggest that this excess may be driven by environmental effects, such as magnetic field compression and relativistic electrons (re-)acceleration due to shocks, or magnetic field amplification via magnetic draping. Consistently, GASP galaxies show $L_\text{R}/L_\text{R}^\text{SFR}>1$ at $\Delta\text{log}_{10}\text{SFR}>0$, i.e. they are radio over-luminous while also lying above the main sequence. This behavior contrasts with the rest of the sample, where $L_\text{R}/L_\text{R}^\text{SFR}>1$ typically corresponds to $\Delta\text{log}_{10}\text{SFR}<0$. This result further supports the presence of an additional contribution to their radio continuum emission that is not directly tied to ongoing SFR, such as shock compression and magnetic draping. Finally, we note that JO147 is significantly more HI-poor than the other two galaxies. This result is consistent with the scenario described in \citet[][]{Ignesti_2026}, in which , despite experiencing strong ram pressure, its low-Mach number motion prevents the formation of a protective magnetic drape, leading to rapid gas loss into the ICM.

\subsection{Cluster environment vs pre-processing}
\label{stack-disc}
The stacking analysis shown in Figure \ref{stack} indicates that cluster galaxies have, overall, $\sim0.7$dex less HI mass than the field represented by xGASS. This holds for both the full sample and the star-forming galaxies subsample, and is consistent with previous results from \citet{Luber_2022}. The deficiency shows a trend with the cluster centric distance, starting at $0.4-0.5$ dex at $R_{200}$, and increases within $R_{500}$, where, overall, cluster galaxies are an order of magnitude more HI-poor than their field counterparts. The large statistics made accessible by the stacking technique has allowed us to sample the cluster volume within $R_{200}$ with enough bins to asses the existence of a gradient in HI mass with the cluster-centric distance.  

The HI-deficiency we observe at $R_{200}$ could likely reflect a combination of contamination from HI-poor galaxies which are moving toward their orbit apocenter \citep[i.e, back-splash galaxies, see][]{Solanes_2001}, which would represent $\sim50\%$ of galaxies at $R>R_{200}$ \citep[][]{Haggar_2020}, and pre-processing, where galaxies enter the cluster having already lost a significant fraction of their gas \citep[][]{Yoon_2025}. Furthermore, the increasing HI deficit within $R_{500}$ indicates that the cluster environment further exacerbate the gas loss, that is in line with the previous studies \citep{Giovannelli_1985,Chung_2007, Deb_2023, Reynolds_2022}. We note, however, that the innermost cluster-centric bin is necessarily contaminated by galaxies projected at small radii but located at larger physical distances, which may retain a higher HI content as they are less affected by the cluster. As a result, the true radial trend is likely steeper than inferred by our analysis. We further note that the HI-selected sample only marginally catches the overall HI deficiency, as their distribution broadly follows that of xGASS. This highlights the importance of stacking techniques to robustly capture the impact of the cluster environment on galaxy gas content. 

\section{Conclusions}
We presented a novel analysis of star-forming galaxies in three nearby clusters making use of the full potential of the MeerKAT radio telescope that permitted us  to study both neutral and nonthermal ISM. We have composed a sample of 61 galaxies selected on the basis of their HI content and radio continuum emission. Such selection intercepts galaxies at the earliest stages of the stripping, i.e. highly asymmetrical systems in which the star formation has been recently enhanced due to the first infall in the cluster, or it has just started declining. This result highlights how full-spectrum MeerKAT studies can be a powerful asset to define samples for next-generation galaxy evolution studies. 

The emerging scenario is that, in the early stripping phases, cluster galaxies are typically found to be HI-deficient ($\Delta\text{log}_{10}M_\text{HI}<0$) but more star forming ($\Delta\text{log}_{10}\text{SFR}>0$), possibly due to gas compression stimulating the SFR. The compression would affect the radio continuum luminosity as well, resulting in a temporary enhancement. Then galaxies move toward a phase in which they are both HI and SFR deficient, but the timescale associated with their SFR evolution are necessarily larger than $\sim10^7$ yr, i.e. the synchrotron cooling time, so that the radio emission is briefly sustained by the old relativistic electrons permeating the ISM. Hence, the ratio $L_\text{R}/L_\text{R}^\text{SFR}$ can be used to disentangle galaxies observed before, during, and right after the SFR boost.

UW and JF show some tentative differences in phase-space location and asymmetry in the radio continuum and HI emission, suggesting an underlying time sequence where UW galaxies evolve into JF. Radio vs HI asymmetry comparison tentatively suggests that neutral ISM become disturbed earlier than the nonthermal one. In line with outside-in stripping, this result implies that the nonthermal ISM, responsible for the radio continuum emission, is bound to the warm, ionized ISM rather than the neutral phase. However, we note that we detect only massive, and mostly over gas-rich UW, hence we suggest that only massive, baryon-dominated galaxies go through the UW phase, maybe due to hydrodynamical instabilities driven by the external processes.

Finally, stacking analysis revealed that cluster galaxies are, on average, more gas-poor than their field counterpart already at $R_{200}$, and the deficiency further grows within $R_{500}$, where star-forming galaxies have 0.5dex less HI than their field counterpart. The real amplitude of the environmental-induced gas loss cannot be fully grasped by studying only HI-detected samples, that highlights the importance of stacking analysis to explore the underlying, undetected population.

This project has showcased the potential of MeerKAT in studying cluster galaxies. The comparison between neutral and nonthermal ISM properties can reveal fundamental properties of the ISM microphysics, which can be further explored via new, tailored MHD simulation able to fully reproduce the ISM-ICM interaction and the full evolution of the multi-phase ISM. Joint HI-radio continuum studies will be routinely possible with SKA-Mid Band 2 observations \citep[][]{braun2019anticipatedperformancesquarekilometre}, where the increased angular resolution offered by SKA-mid will allow us to localize these effects, rather than being currently limited to global properties. Future HI-radio continuum studies will permit us to intercept galaxies at the initial phase of environmental processing and push forward our understanding of this driver of galaxy evolution. 

\begin{acknowledgements}
We acknowledge the constructive contribute of the Referee that improved the presentation of our work. The MeerKAT telescope is operated by the South African Radio Astronomy Observatory, which is a facility of the National Research Foundation, an agency of the Department of Science and Innovation. Based on observations collected at the European Organization for Astronomical Research in the Southern Hemisphere under ESO programme 196.B-0578. This project has received funding from the European Research Council (ERC) under the European Union's Horizon 2020 research and innovation programme (grant agreement No. 833824). (Part of) the data published here have been reduced using the CARACal pipeline, partially supported by ERC Starting grant number 679627 “FORNAX”, MAECI Grant Number ZA18GR02, DST-NRF Grant Number 113121 as part of the ISARP Joint Research Scheme, and BMBF project 05A17PC2 for D-MeerKAT. AI acknowledges support from the institutional project RVO:67985815 and the project 25-19512L of the Czech Science Foundation. BV and AEL acknowledge support from the INAF GO grant 2023 ``Identifying ram pressure induced unwinding arms in cluster spirals'' (P.I. Vulcani). The Enhancement of the SRT for the study of the Universe at high radio frequencies is financially supported by the National Operative Program (Programma Operativo Nazionale - PON) of the Italian Ministry of University and Research "Research and Innovation 2014-2020", Notice D.D. 424 of 28/02/2018 for the granting of funding aimed at strengthening research infrastructures, in implementation of the Action II.1 – Project Proposal PIR01$\_$00010. This work was carried out thanks to the funding of the Regione Autonoma della Sardegna, ai sensi della Legge Regionale 7 agosto 2007, n.7 "Promozione della Ricerca Scientifica e dell'Innovazione Tecnologica in Sardegna".
AB and LB acknowledge the support of the INAF 2024 Mini-grant Super-MIGHTEE and SPRITZ: paving the road to SKA. CP acknowledges support by the European Research Council under ERC-AdG grant PICOGAL-101019746. NT acknowledge the support by the European Union – NextGenerationEU through the National Recovery and Resilience Plan 2021-2026. Institutional grant of University of Zagreb Faculty of Science ProPubFO-1.1.3.2026. AI thanks the music of Deep Purple for inspiring the preparation of the draft.

\end{acknowledgements}
\bibliographystyle{aa}
\bibliography{sn-bibliography}%
\appendix

\captionsetup{width=\textwidth}
\onecolumn
\section{Full sample}
\begin{figure*}[th!]
\centering %sidecaption %
  \includegraphics[width=\linewidth]{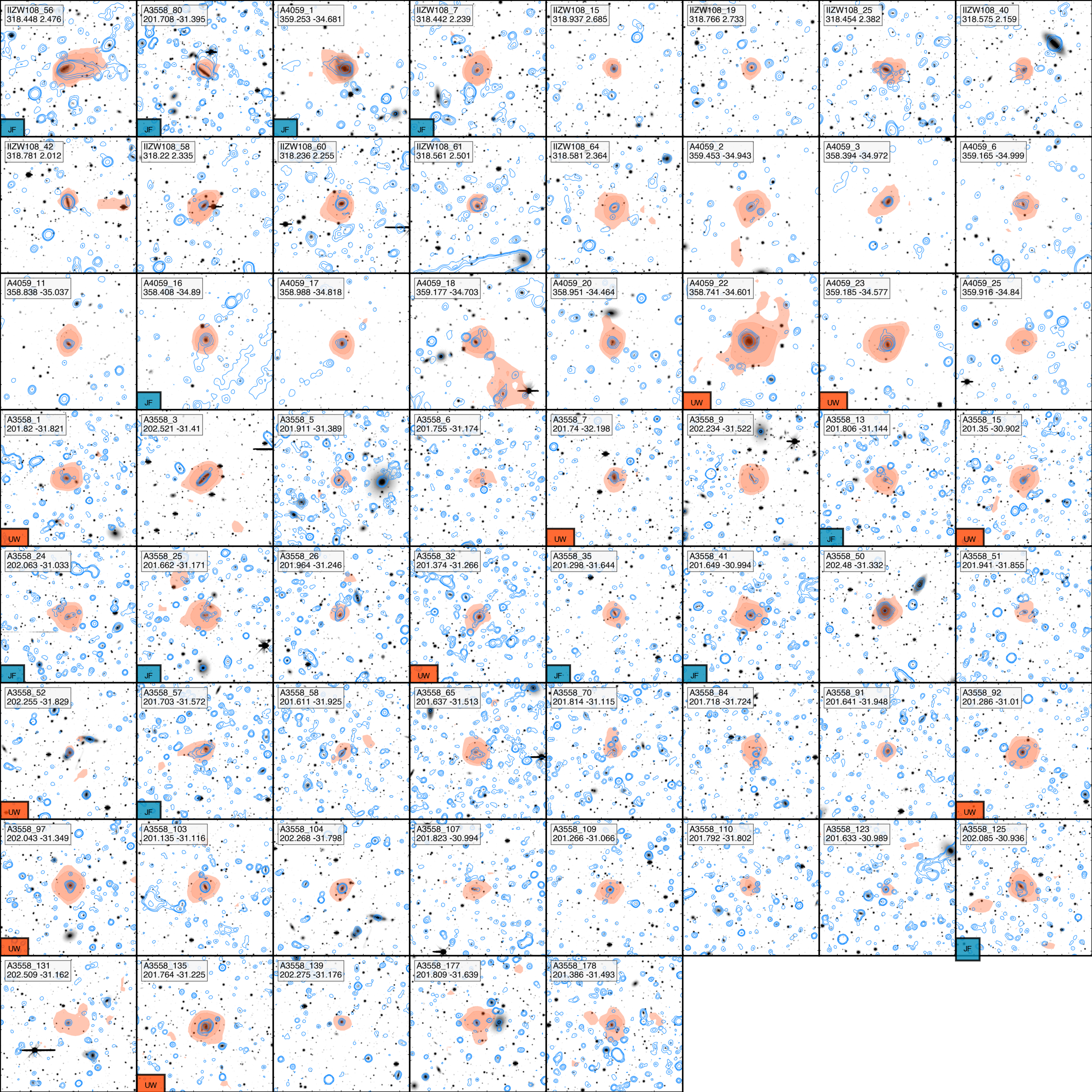}
  \caption{Galaxy sample. For each galaxy we show a 6 arcmin cutout ($\sim360$ kpc at cluster redshift) of Legacy R-band images overlapped with the MeerKAT HI emission (red filled contours) and radio continuum (blue contours) at the 3, 6, 12, 24$\sigma$ levels. In the top panel are reported galaxy ID and optical center J2000 coordinates. UW and JF galaxies are labeled accordingly in the bottom-left corner.}
  \label{mosaic}
\end{figure*}

We report here the combined images for the full sample (Figure \ref{mosaic}) and the detailed information (Table \ref{long_tab}).
\newpage
\begin{longtable}{@{\extracolsep{5pt}}cccccccccc}

\toprule
\toprule
Cluster	&ID&	$M_*$	&	SFR			&	$M_{\text{HI}}$			&	$L_R$			&	$A_{\text{HI}}$	&	$A_{\text{R}}$	&	$C_{\text{PS}}$	&	$C_{\text{O}}$	\\
&& [log$_{10}(M_{\odot})$] & [log$_{10}(M_{\odot}~\text{yr}^{-1})$] & [log$_{10}(M_{\odot})$] & [log$_{10}(\text{erg s}^{-1})$] &&&&\\
\midrule
\endhead

\bottomrule
\endfoot
A4059	&	1	&	11.47	$\pm$	0.1	&	1.17	$\pm$	0.14	&	9.58	$\pm$	0.02	&	29.83	$\pm$	0.04	&	0.68	&	0.73	&	3	&	2	\\
A3558	&	80	&	11.05	$\pm$	0.12	&	1.02	$\pm$	0.17	&	9.07	$\pm$	0.02	&	29.85	$\pm$	0.02	&	0.67	&	0.94	&	1	&	2	\\
IIZW108	&	56	&	10.84	$\pm$	0.12	&	0.96	$\pm$	0.41	&	9.73	$\pm$	0.02	&	29.73	$\pm$	0.02	&	0.79	&	0.94	&	1	&	2	\\
\midrule
IIZW108	&	7	&	9.35	$\pm$	0.18	&	0	$\pm$	0.28	&	9.47	$\pm$	0.02	&	28.33	$\pm$	0.03	&	0.6	&	0.68	&	1	&	2	\\
IIZW108	&	15	&	10.28	$\pm$	0.13	&	0.26	$\pm$	0.47	&	8.9	$\pm$	0.02	&	27.82	$\pm$	0.07	&	0.58	&	0.72	&	1	&	0	\\
IIZW108	&	19	&	9.77	$\pm$	0.14	&	-0.07	$\pm$	0.61	&	8.78	$\pm$	0.02	&	28.29	$\pm$	0.03	&	0.64	&	0.48	&	1	&	0	\\
IIZW108	&	25	&	11.11	$\pm$	0.13	&	0.29	$\pm$	0.88	&	9.19	$\pm$	0.02	&	29.04	$\pm$	0.02	&	0.73	&	0.85	&	3	&	0	\\
IIZW108	&	40	&	10.06	$\pm$	0.13	&	0.22	$\pm$	0.44	&	9.01	$\pm$	0.02	&	28.08	$\pm$	0.06	&	0.58	&	0.88	&	4	&	0	\\
IIZW108	&	42	&	11.11	$\pm$	0.14	&	0.69	$\pm$	0.93	&	9.11	$\pm$	0.02	&	29.69	$\pm$	0.02	&	0.96	&	0.72	&	1	&	0	\\
IIZW108	&	58	&	10.92	$\pm$	0.14	&	0.3	$\pm$	0.71	&	9.44	$\pm$	0.02	&	28.29	$\pm$	0.03	&	0.31	&	0.29	&	1	&	0	\\
IIZW108	&	60	&	10.62	$\pm$	0.12	&	0.61	$\pm$	0.5	&	9.42	$\pm$	0.02	&	28.95	$\pm$	0.02	&	0.36	&	0.46	&	1	&	0	\\
IIZW108	&	61	&	10.09	$\pm$	0.14	&	0.41	$\pm$	0.38	&	8.89	$\pm$	0.02	&	28.77	$\pm$	0.02	&	0.68	&	0.79	&	1	&	0	\\
IIZW108	&	64	&	9.68	$\pm$	0.15	&	-0.36	$\pm$	0.86	&	9.39	$\pm$	0.02	&	28.37	$\pm$	0.03	&	0.53	&	0.14	&	3	&	0	\\
\midrule
A4059	&	2	&	9.37	$\pm$	0.17	&	-0.11	$\pm$	0.2	&	9.53	$\pm$	0.02	&	28.45	$\pm$	0.04	&	0.7	&	0.69	&	3	&	0	\\
A4059	&	3	&	10.91	$\pm$	0.11	&	0.46	$\pm$	0.64	&	9.28	$\pm$	0.02	&	28.99	$\pm$	0.03	&	0.74	&	0.24	&	2	&	0	\\
A4059	&	6	&	9.58	$\pm$	0.19	&	0.08	$\pm$	0.44	&	9.2	$\pm$	0.02	&	28.62	$\pm$	0.03	&	0.7	&	0.41	&	1	&	0	\\
A4059	&	11	&	10.06	$\pm$	0.14	&	0.45	$\pm$	0.4	&	9.29	$\pm$	0.02	&	28.64	$\pm$	0.03	&	0.65	&	0.65	&	4	&	0	\\
A4059	&	16	&	9.93	$\pm$	0.17	&	0.67	$\pm$	0.3	&	9.26	$\pm$	0.02	&	29.18	$\pm$	0.03	&	0.67	&	0.83	&	1	&	2	\\
A4059	&	17	&	10.05	$\pm$	0.16	&	0.57	$\pm$	0.38	&	9.18	$\pm$	0.02	&	28.67	$\pm$	0.03	&	0.73	&	0.57	&	1	&	0	\\
A4059	&	18	&	10.12	$\pm$	0.14	&	0.41	$\pm$	0.38	&	9.76	$\pm$	0.02	&	28.94	$\pm$	0.03	&	0.88	&	0.84	&	5	&	0	\\
A4059	&	20	&	9.99	$\pm$	0.14	&	0.45	$\pm$	0.36	&	9.36	$\pm$	0.02	&	28.7	$\pm$	0.03	&	0.56	&	0.4	&	3	&	0	\\
A4059	&	22	&	11.4	$\pm$	0.12	&	1.19	$\pm$	0.49	&	10.13	$\pm$	0.02	&	29.65	$\pm$	0.02	&	0.6	&	0.44	&	1	&	1	\\
A4059	&	23	&	10.62	$\pm$	0.13	&	0.81	$\pm$	0.43	&	9.86	$\pm$	0.02	&	29.42	$\pm$	0.02	&	0.43	&	0.71	&	3	&	1	\\
A4059	&	25	&	9.1	$\pm$	0.22	&	-0.12	$\pm$	0.19	&	9.19	$\pm$	0.02	&	27.82	$\pm$	0.09	&	0.49	&	0.68	&	1	&	0	\\
\midrule
A3558	&	1	&	10.14	$\pm$	0.13	&	0.1	$\pm$	0.2	&	9.58	$\pm$	0.02	&	28.36	$\pm$	0.03	&	0.59	&	0.65	&	3	&	1	\\
A3558	&	3	&	11.26	$\pm$	0.12	&	1.15	$\pm$	0.45	&	9.79	$\pm$	0.02	&	29.75	$\pm$	0.02	&	0.6	&	0.63	&	1	&	0	\\
A3558	&	5	&	9.98	$\pm$	0.12	&	-0.03	$\pm$	0.2	&	9.04	$\pm$	0.02	&	28.65	$\pm$	0.02	&	0.83	&	0.79	&	5	&	0	\\
A3558	&	6	&	8.75	$\pm$	0.17	&	-0.86	$\pm$	0.24	&	8.95	$\pm$	0.02	&	27.16	$\pm$	0.09	&	0.79	&	0.66	&	1	&	0	\\
A3558	&	7	&	10.24	$\pm$	0.14	&	0.49	$\pm$	0.21	&	9.2	$\pm$	0.02	&	28.89	$\pm$	0.03	&	0.65	&	0.51	&	1	&	1	\\
A3558	&	9	&	9.77	$\pm$	0.13	&	-0.27	$\pm$	0.2	&	9.59	$\pm$	0.02	&	28.09	$\pm$	0.06	&	0.66	&	0.74	&	1	&	0	\\
A3558	&	13	&	9.46	$\pm$	0.13	&	-0.44	$\pm$	0.23	&	9.46	$\pm$	0.02	&	27.56	$\pm$	0.06	&	0.27	&	0.87	&	4	&	2	\\
A3558	&	15	&	10.24	$\pm$	0.11	&	-0.33	$\pm$	0.49	&	9.47	$\pm$	0.02	&	28.03	$\pm$	0.04	&	0.67	&	0.83	&	1	&	1	\\
A3558	&	24	&	9.47	$\pm$	0.15	&	-0.47	$\pm$	0.3	&	9.46	$\pm$	0.02	&	27.69	$\pm$	0.06	&	0.48	&	0.83	&	4	&	2	\\
A3558	&	25	&	9.81	$\pm$	0.14	&	-0.04	$\pm$	0.22	&	9.73	$\pm$	0.02	&	28.32	$\pm$	0.03	&	0.65	&	0.67	&	4	&	2	\\
A3558	&	26	&	9.62	$\pm$	0.19	&	0.41	$\pm$	0.27	&	9.02	$\pm$	0.02	&	29.09	$\pm$	0.02	&	0.82	&	0.9	&	4	&	0	\\
A3558	&	32	&	10.43	$\pm$	0.11	&	0.06	$\pm$	0.22	&	9.13	$\pm$	0.02	&	28.75	$\pm$	0.02	&	0.33	&	0.57	&	4	&	1	\\
A3558	&	35	&	9.84	$\pm$	0.13	&	-0.27	$\pm$	0.24	&	9.27	$\pm$	0.02	&	28.48	$\pm$	0.03	&	0.65	&	0.64	&	4	&	2	\\
A3558	&	41	&	9.84	$\pm$	0.17	&	0.34	$\pm$	0.19	&	9.57	$\pm$	0.02	&	28.67	$\pm$	0.02	&	0.97	&	0.73	&	4	&	2	\\
A3558	&	50	&	11.34	$\pm$	0.09	&	0.36	$\pm$	0.18	&	9.5	$\pm$	0.02	&	29.42	$\pm$	0.02	&	0.69	&	0.61	&	4	&	0	\\
A3558	&	51	&	8.98	$\pm$	0.2	&	-0.18	$\pm$	0.31	&	8.95	$\pm$	0.02	&	27.49	$\pm$	0.07	&	0.84	&	0.6	&	4	&	0	\\
A3558	&	52	&	10.51	$\pm$	0.12	&	0.35	$\pm$	0.48	&	9	$\pm$	0.02	&	28.4	$\pm$	0.03	&	1	&	0.45	&	4	&	1	\\
A3558	&	57	&	10.21	$\pm$	0.16	&	0.88	$\pm$	0.15	&	9.24	$\pm$	0.02	&	29.19	$\pm$	0.02	&	0.72	&	0.79	&	4	&	2	\\
A3558	&	58	&	10.31	$\pm$	0.1	&	-0.66	$\pm$	0.27	&	9.05	$\pm$	0.02	&	27.8	$\pm$	0.04	&	0.85	&	0.83	&	4	&	0	\\
A3558	&	65	&	9.49	$\pm$	0.13	&	-0.41	$\pm$	0.22	&	9.26	$\pm$	0.02	&	28	$\pm$	0.03	&	0.55	&	0.75	&	4	&	0	\\
A3558	&	70	&	9.27	$\pm$	0.14	&	-0.97	$\pm$	0.35	&	8.99	$\pm$	0.02	&	28.23	$\pm$	0.03	&	0.88	&	0.85	&	4	&	0	\\
A3558	&	84	&	8.84	$\pm$	0.18	&	-0.83	$\pm$	0.26	&	9.26	$\pm$	0.02	&	28.03	$\pm$	0.03	&	0.81	&	0.61	&	1	&	0	\\
A3558	&	91	&	9.98	$\pm$	0.14	&	0.07	$\pm$	0.22	&	8.75	$\pm$	0.02	&	28.51	$\pm$	0.02	&	0.6	&	0.39	&	1	&	0	\\
A3558	&	92	&	10.17	$\pm$	0.17	&	0.82	$\pm$	0.49	&	9.63	$\pm$	0.02	&	28.85	$\pm$	0.02	&	0.31	&	0.39	&	1	&	1	\\
A3558	&	97	&	10.46	$\pm$	0.12	&	0.45	$\pm$	0.18	&	9.76	$\pm$	0.02	&	28.68	$\pm$	0.02	&	0.68	&	0.52	&	1	&	1	\\
A3558	&	103	&	10.47	$\pm$	0.15	&	0.82	$\pm$	0.39	&	9.46	$\pm$	0.02	&	29.27	$\pm$	0.02	&	0.57	&	0.71	&	1	&	2	\\
A3558	&	104	&	10.17	$\pm$	0.14	&	0.72	$\pm$	0.32	&	9	$\pm$	0.02	&	28.94	$\pm$	0.02	&	0.53	&	0.59	&	1	&	0	\\
A3558	&	107	&	9.68	$\pm$	0.12	&	-0.67	$\pm$	0.26	&	9.04	$\pm$	0.02	&	27.35	$\pm$	0.08	&	0.52	&	0.7	&	1	&	0	\\
A3558	&	109	&	9.67	$\pm$	0.16	&	0.78	$\pm$	0.36	&	9.16	$\pm$	0.02	&	28.34	$\pm$	0.03	&	0.44	&	0.62	&	1	&	0	\\
A3558	&	110	&	9.06	$\pm$	0.15	&	-0.84	$\pm$	0.25	&	8.74	$\pm$	0.02	&	27.24	$\pm$	0.08	&	0.71	&	0.68	&	3	&	0	\\
A3558	&	123	&	9.3	$\pm$	0.15	&	-0.4	$\pm$	0.25	&	9.05	$\pm$	0.02	&	27.97	$\pm$	0.03	&	0.64	&	0.74	&	1	&	0	\\
A3558	&	125	&	10.59	$\pm$	0.1	&	-0.06	$\pm$	0.18	&	9.76	$\pm$	0.02	&	28.68	$\pm$	0.02	&	1	&	0.69	&	1	&	2	\\
A3558	&	131	&	9.7	$\pm$	0.16	&	0.15	$\pm$	0.29	&	9.51	$\pm$	0.02	&	28.21	$\pm$	0.04	&	0.79	&	0.67	&	1	&	0	\\
A3558	&	135	&	10.94	$\pm$	0.1	&	0.42	$\pm$	0.19	&	9.78	$\pm$	0.02	&	29.1	$\pm$	0.02	&	0.99	&	0.67	&	3	&	1	\\
A3558	&	139	&	9.95	$\pm$	0.13	&	-0.09	$\pm$	0.21	&	8.81	$\pm$	0.02	&	28.02	$\pm$	0.04	&	0.9	&	0.46	&	3	&	0	\\
A3558	&	177	&	10.35	$\pm$	0.1	&	-0.52	$\pm$	0.24	&	9.59	$\pm$	0.02	&	27.79	$\pm$	0.04	&	0.73	&	0.62	&	3	&	0	\\
A3558	&	178	&	10.4	$\pm$	0.11	&	-0.23	$\pm$	0.23	&	9.52	$\pm$	0.02	&	28.49	$\pm$	0.02	&	0.73	&	0.45	&	1	&	0	\\

%\end{tabular}
%\end{table*}
\bottomrule
\caption{\label{long_tab} Galaxy sample properties. From left to right: hosting cluster name; HI detection ID; Stellar mass; Star formation rate; HI mass; Radio continuum luminosity at 1.4 GHz; HI asymmetry; Radio continuum asymmetry; Phase-space classification (1= undetermined, 2= first infallers, 3= recent infallers, 4= intermediate infallers, 5= ancient infallers); Optical feature classification (0= unclassified, 1= unwinding candidate, 2= jellyfish candidate).}

\end{longtable}

\section{Stacked spectra}
\begin{figure*}[h!]
    \includegraphics[width=.5\linewidth]{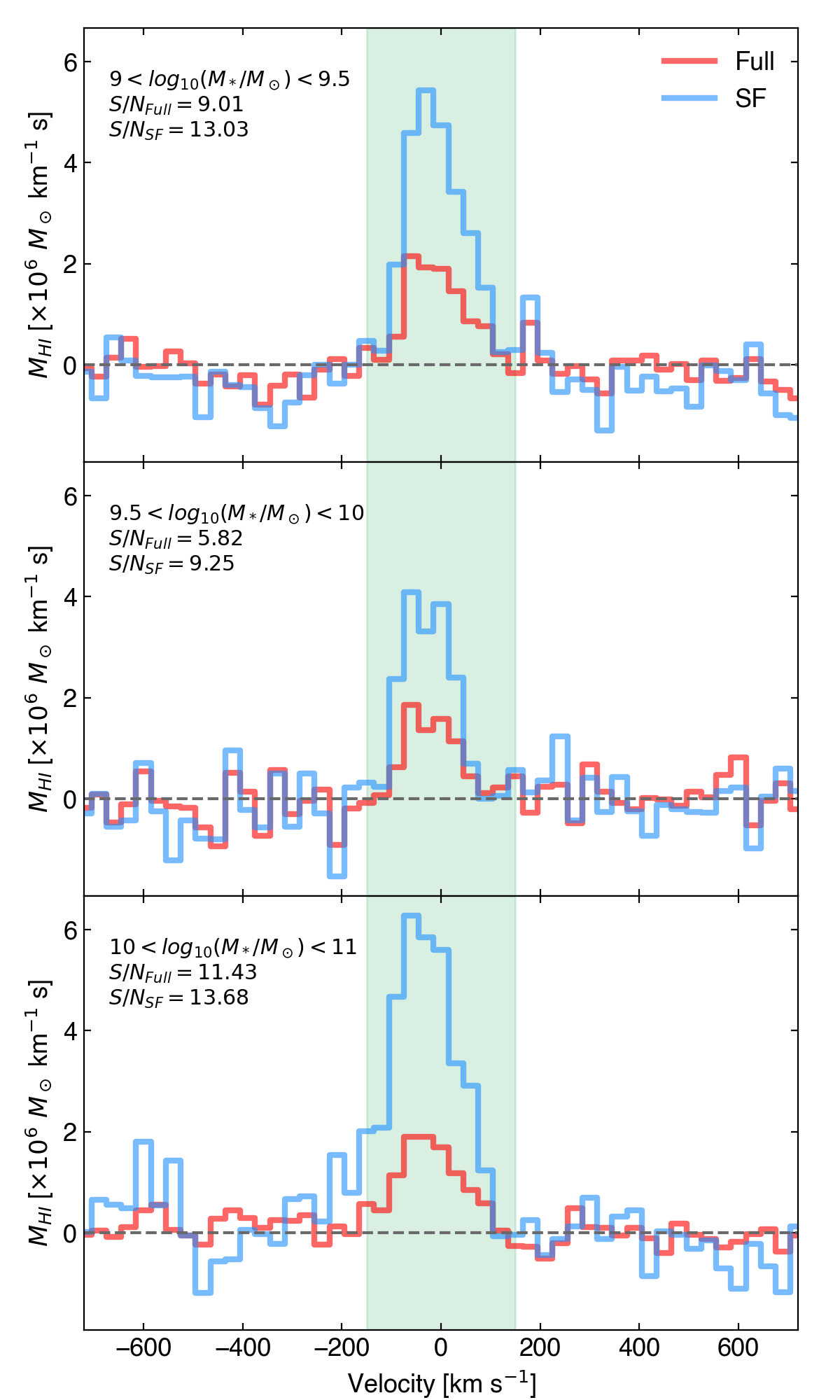}
    \includegraphics[width=.5\linewidth]{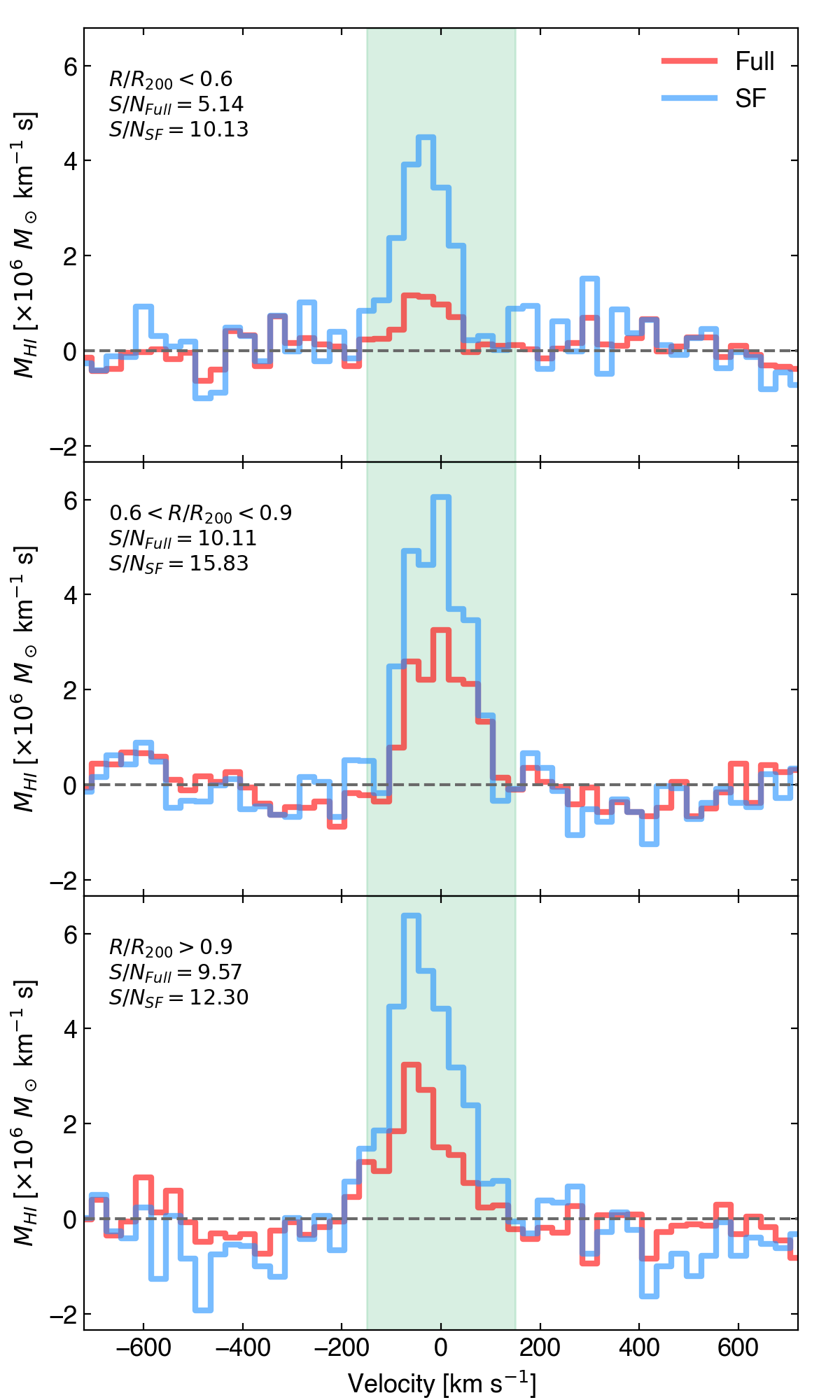}
    \caption{\label{stack_spectra} Stacked spectra for stellar mass (left) and cluster-centric distance (right) derived from the full (red) and star-forming (SF, blue) samples. The green vertical band indicates the integration window used to compute the average HI mass. For reference, the dashed black line points the zeroth level. Bin size and signal-to-noise ratios of each spectra are reported in the top-left corner.}
\end{figure*}

In Figure \ref{stack_spectra} are presented the stacked spectra from which we derived the average HI mass reported in Table \ref{tab_stack} and Figure \ref{stack}.
\end{document}